\documentclass{nature_modified}
\usepackage{hyperref}
\usepackage{graphicx}
\usepackage{multibib}
\usepackage{lscape}
\usepackage{xcolor}
\usepackage{float}
\usepackage{amsmath}
\usepackage{amssymb}
\usepackage{longtable}
\usepackage{caption}
\newcites{Methods}{References}

\spacing{1}

\title{Transit detection of the long-period volatile-rich super-Earth \hbox{$\nu^2$ Lupi d} with \textit{CHEOPS}}

\author{Laetitia Delrez$^{1,2,3}$,
David Ehrenreich$^{3}$,
Yann Alibert$^{4}$,
Andrea Bonfanti$^{5}$,
Luca Borsato$^{6}$,
Luca Fossati$^{5}$,
Matthew J. Hooton$^{4}$,
Sergio Hoyer$^{7}$,
Francisco J. Pozuelos$^{1,2}$,
Sébastien Salmon$^{2,3}$,
Sophia Sulis$^{7}$,
Thomas G. Wilson$^{8}$,
Vardan Adibekyan$^{9,10}$,
Vincent Bourrier$^{3}$,
Alexis Brandeker$^{11}$,
Sébastien Charnoz$^{12}$,
Adrien Deline$^{3}$,
Pascal Guterman$^{7,13}$,
Jonas Haldemann$^{4}$, 
Nathan Hara$^{3}$,
Mahmoudreza Oshagh$^{14,15}$,
Sergio G. Sousa$^{9}$,
Valérie Van Grootel$^{2}$,
Roi Alonso$^{14,15}$,
Guillem Anglada Escudé$^{16,17}$,
Tamás Bárczy$^{18}$,
David Barrado$^{19}$,
Susana C. C. Barros$^{9,10}$,
Wolfgang Baumjohann$^{5}$,
Mathias Beck$^{3}$,
Anja Bekkelien$^{3}$,
Willy Benz$^{4,20}$,
Nicolas Billot$^{3}$,
Xavier Bonfils$^{21}$,
Christopher Broeg$^{4}$,
Juan Cabrera$^{22}$,
Andrew Collier Cameron$^{8}$,
Melvyn B. Davies$^{23}$,
Magali Deleuil$^{7}$,
Jean-Baptiste Delisle$^{3}$,
Olivier D. S. Demangeon$^{9,10}$,
Brice-Olivier Demory$^{20}$,
Anders Erikson$^{22}$,
Andrea Fortier$^{4}$,
Malcolm Fridlund$^{24,25}$,
David Futyan$^{3}$,
Davide Gandolfi$^{26}$,
Antonio Garcia Muñoz$^{27}$,
Michaël Gillon$^{1}$,
Manuel Guedel$^{28}$,
Kevin Heng$^{20}$,
László Kiss$^{29,30,31}$,
Jacques Laskar$^{32}$,
Alain Lecavelier des Etangs$^{33}$,
Monika Lendl$^{3,5}$,
Christophe Lovis$^{3}$,
Pierre F. L. Maxted$^{34}$,
Valerio Nascimbeni$^{6}$,
Göran Olofsson$^{11}$,
Hugh P. Osborn$^{35,36}$,
Isabella Pagano$^{37}$,
Enric Pallé$^{14,15}$,
Giampaolo Piotto$^{6,38}$,
Don Pollacco$^{39}$,
Didier Queloz$^{3,40}$,
Heike Rauer$^{22,27,41}$,
Roberto Ragazzoni$^{6}$,
Ignasi Ribas$^{16,17}$,
Nuno C. Santos$^{9,10}$,
Gaetano Scandariato$^{37}$,
Damien Ségransan$^{3}$,
Attila E. Simon$^{4}$,
Alexis M. S. Smith$^{22}$,
Manfred Steller$^{5}$,
Gyula M. Szab{\'o}$^{42,43}$,
Nicolas Thomas$^{4}$,
Stéphane Udry$^{3}$,
and Nicholas A. Walton$^{44}$
}

\begin{document}

\maketitle

\begin{affiliations}
\begin{footnotesize}

%
\item Astrobiology Research Unit, Universit{\'e} de Li{\`e}ge, All\'ee du 6 Ao{\^u}t 19C, 4000 Li{\`e}ge, Belgium
%
\item Space sciences, Technologies and Astrophysics Research (STAR) Institute, Universit{\'e} de Li{\`e}ge, All{\'e}e du 6 Ao{\^u}t 19C, 4000 Li{\`e}ge, Belgium 
%
\item Observatoire de Gen\`eve, Universit\'e de Gen\`eve, Chemin Pegasi 51, 1290 Versoix, Switzerland
%
\item Physikalisches Institut, University of Bern, Gesellsschaftstrasse 6, 3012 Bern, Switzerland
%
\item Space Research Institute, Austrian Academy of Sciences, Schmiedlstrasse 6, A-8042 Graz, Austria
%
\item INAF, Osservatorio Astronomico di Padova, Vicolo dell'Osservatorio 5, 35122 Padova, Italy
%
\item Aix Marseille Univ, CNRS, CNES, LAM, Marseille, France
%
\item Centre for Exoplanet Science, SUPA School of Physics and Astronomy, University of St Andrews, North Haugh, St Andrews KY16 9SS, UK
%
\item Instituto de Astrof\'isica e Ci\^encias do Espa\c{c}o, Universidade do Porto, CAUP, Rua das Estrelas, 4150-762 Porto, Portugal
%
\item Departamento de F\'isica e Astronomia, Faculdade de Ci\^encias, Universidade do Porto, Rua do Campo Alegre, 4169-007 Porto, Portugal
%
\item Department of Astronomy, Stockholm University, AlbaNova University Center, 10691 Stockholm, Sweden
%
\item Institut de Physique du Globe de Paris (IPGP), 1 rue Jussieu, 75005 Paris, France
%
\item Division Technique INSU, BP 330, 83507 La Seyne cedex, France
%
\item Instituto de Astrofísica de Canarias, 38200 La Laguna, Tenerife, Spain
%
\item Departamento de Astrofísica, Universidad de La Laguna, 38206 La Laguna, Tenerife, Spain
%
\item Institut de Ciències de l’Espai (ICE, CSIC), Campus UAB, Can Magrans s/n, 08193 Bellaterra, Spain
%
\item Institut d’Estudis Espacials de Catalunya (IEEC), 08034 Barcelona, Spain
%
\item Admatis, Miskok, Hungary
%
\item Depto. de Astrofísica, Centro de Astrobiologia (CSIC-INTA), ESAC campus, 28692 Villanueva de la Cãda (Madrid), Spain
%
\item Center for Space and Habitability, Gesellsschaftstrasse 6, 3012 Bern, Switzerland
%
\item Université Grenoble Alpes, CNRS, IPAG, 38000 Grenoble, France
%
\item Institute of Planetary Research, German Aerospace Center (DLR), Rutherfordstrasse 2, 12489 Berlin, Germany
%
\item Lund Observatory, Dept. of Astronomy and Theoretical Physics, Lund University, Box 43, 22100 Lund, Sweden
%
\item Leiden Observatory, University of Leiden, PO Box 9513, 2300 RA Leiden, The Netherlands
%
\item Department of Space, Earth and Environment, Chalmers University of Technology, Onsala Space Observatory, 43992 Onsala, Sweden
%
\item INAF, Osservatorio Astrofisico di Torino, via Osservatorio 20, 10025 Pino Torinese, Italy
%
\item Center for Astronomy and Astrophysics, Technical University Berlin, Hardenberstrasse 36, 10623 Berlin, Germany
%
\item University of Vienna, Department of Astrophysics, Türkenschanzstrasse 17, 1180 Vienna, Austria
%
\item Konkoly Observatory, Research Centre for Astronomy and Earth Sciences, 1121 Budapest, Konkoly Thege Miklós út 15-17, Hungary
%
\item ELTE E\"otv\"os Lor\'and University, Institute of Physics, P\'azm\'any P\'eter s\'et\'any 1/A, 1117 Budapest, Hungary
%
\item Sydney Institute for Astronomy, School of Physics A29, University of Sydney, NSW 2006, Australia
%
\item IMCCE, UMR8028 CNRS, Observatoire de Paris, PSL Univ., Sorbonne Univ., 77 av. Denfert-Rochereau, 75014 Paris, France
%
\item Institut d’astrophysique de Paris, UMR7095 CNRS, Université Pierre \& Marie Curie, 98bis blvd. Arago, 75014 Paris, France
%
\item Astrophysics Group, Keele University, Staffordshire, ST5 5BG, United Kingdom
%
\item NCCR/PlanetS, Centre for Space \& Habitability, University of Bern, Bern, Switzerland
%
\item Department of Physics and Kavli Institute for Astrophysics and Space Research, Massachusetts Institute of Technology, Cambridge, MA 02139, USA
%
\item INAF, Osservatorio Astrofisico di Catania, Via S. Sofia 78, 95123 Catania, Italy
%
\item Dipartimento di Fisica e Astronomia "Galileo Galilei", Università degli Studi di Padova, Vicolo dell'Osservatorio 3, 35122 Padova, Italy
%
\item Department of Physics, University of Warwick, Gibbet Hill Road, Coventry CV4 7AL, United Kingdom
%
\item Astrophysics Group, Cavendish Laboratory, J.J. Thomson Avenue, Cambridge CB3 0HE, United Kingdom
%
\item Institut für Geologische Wissenschaften, Freie Universität Berlin, 12249 Berlin, Germany
%
\item ELTE Eötvös Loránd University, Gothard Astrophysical Observatory, 9700 Szombathely, Szent Imre h. u. 112, Hungary
%
\item MTA-ELTE Exoplanet Research Group, 9700 Szombathely, Szent Imre h. u. 112, Hungary
%
\item Institute of Astronomy, University of Cambridge, Madingley Road, Cambridge, CB3 0HA, United Kingdom

\end{footnotesize}
\end{affiliations}


\begin{abstract}
Exoplanets transiting bright nearby stars are key objects for advancing our knowledge of planetary formation and evolution. The wealth of photons from the host star gives detailed access to the atmospheric, interior, and orbital properties of the planetary companions. \hbox{$\nu^2$ Lupi} (HD 136352) is a naked-eye ($V = 5.78$) Sun-like star that was discovered to host three low-mass planets with orbital periods of 11.6, 27.6, and 107.6 days via radial velocity monitoring\cite{2019A&A...622A..37U}. The two inner planets (b and c) were recently found to transit\cite{2020AJ....160..129K}, prompting a photometric follow-up by the brand-new \textit{CHaracterising ExOPlanets Satellite (CHEOPS)}. Here, we report that the outer planet d is also transiting, and measure its radius and mass to be \hbox{$2.56\pm0.09$ $R_{\oplus}$} and \hbox{$8.82\pm0.94$ $M_{\oplus}$}, respectively. With its bright Sun-like star, long period, and mild irradiation ($\sim$5.7 times the irradiation of Earth), $\nu^2$ Lupi d unlocks a completely new region in the parameter space of exoplanets amenable to detailed characterization. We refine the properties of all three planets: planet b likely has a rocky mostly dry composition, while planets c and d seem to have retained small hydrogen-helium envelopes and a possibly large water fraction. This diversity of planetary compositions makes the \hbox{$\nu^2$ Lupi} system an excellent laboratory for testing formation and evolution models of low-mass planets.
\end{abstract}

\textit{CHEOPS}\cite{willy} is the new European mission dedicated to the study of known exoplanets around bright stars ($V\leq$ 12). Unlike previous exoplanet detection missions, like \textit{CoRoT}\cite{2006ESASP1306...33B}, \textit{Kepler}\cite{2010Sci...327..977B}, and \textit{TESS}\cite{2015JATIS...1a4003R}, \textit{CHEOPS} is a follow-up mission, designed to collect ultra-high precision photometry of a single star at a time. For this purpose, it relies on a 30 cm effective aperture telescope, equipped with a single frame-transfer back-illuminated CCD detector providing a broad 330-1100 nm bandpass\cite{2020A&A...635A..22D}. \textit{CHEOPS} was launched on 18 December 2019 into a 700 km altitude Sun-synchronous dusk-dawn orbit and started routine science operations in April 2020. For very bright stars ($V\sim6$), \textit{CHEOPS} demonstrated that it can achieve an outstanding photometric precision of about 10 parts per million (ppm) per 1-hour intervals\cite{W189}.

$\nu^2$ Lupi is one of the first scientific targets observed by \textit{CHEOPS}. This system of three low-mass planets orbiting one of the closest (14.7 parsecs) G-type main-sequence stars was discovered using radial velocities (RVs) obtained with the HARPS spectrograph\cite{2019A&A...622A..37U}. It was then observed by \textit{TESS} during Sector 12 of its primary mission (21 May-18 June 2019), which revealed that the two inner planets are transiting\cite{2020AJ....160..129K}. These 28-day long observations did not cover any inferior conjunction of the outer planet d. However, the transiting configuration of the two inner planets increased the probability that it is also transiting, to about 20\% for typical mutual inclinations of $\sim$1 deg\cite{2020AJ....160..129K}. \hbox{$\nu^2$ Lupi} is one of only three naked-eye stars known to host several transiting planets, the other two being HD 219134\cite{2015A&A...584A..72M,2017NatAs...1E..56G} and HR 858\cite{2019ApJ...881L..19V}. The multi-transiting nature of these systems, combined with the brightness of their star, make them targets of paramount importance for comparative exoplanetology studies. The primary objective of our follow-up of $\nu^2$ Lupi with \textit{CHEOPS} was to refine the properties of the two inner planets, most notably their radii but also their ephemerides, since being able to predict precise transit times is essential to enable follow-up observations with heavily-subscribed facilities\cite{2020AJ....159..219D}. 

Six observation runs (visits) were obtained with \textit{CHEOPS} between 4 April and 6 July 2020 (Supplementary Table 1), targeting four transits of planet b and three of planet c (the last visit contained one transit of each planet). The data were processed with the \textit{CHEOPS} data reduction pipeline\cite{2020A&A...635A..24H} (see Methods) and the resulting individual light curves are shown in Supplementary Figure 1. During the fifth visit (8-9 June 2020), we serendipitously detected a $\sim$500 ppm transit-like flux drop which started during the targeted transit of planet c, and lasted for the rest of the visit (Supplementary Figure 1, bottom left panel). We carefully checked the data for systematics and found this signal to be very robust (see Methods). Furthermore, the star does not show any comparable photometric variability in neither the \textit{CHEOPS} nor \textit{TESS} data (Supplementary \hbox{Figure 4)}. This flux drop occurred at 1.3$\sigma$ of an inferior conjunction of \hbox{$\nu^2$ Lupi d}, as predicted by the RV orbital solution of ref. \cite{2019A&A...622A..37U}, and we show in Methods that it most likely originates from a transit of this outer planet. This makes $\nu^2$ Lupi d the first long-period ($>$100 days) planet detected to transit a naked-eye star (Figure~\ref{fig:diagrams}a). 

To determine the system parameters, we performed a global analysis of our six \textit{CHEOPS} light curves together with other available photometric and RV data. This includes first a custom \textit{TESS} light curve, which we extracted from the target pixel files following ref. \cite{2020AJ....160..129K}, to conduct a careful treatment of instrumental systematics (see Methods and Supplementary Figure 4). We also included 246 previously published\cite{2019A&A...622A..37U,2020AJ....160..129K} RV measurements obtained with HARPS between May 2004 and August 2017 (see Methods and Supplementary Figure 6).

\begin{figure*}
\centering
\includegraphics[width=0.8\hsize]{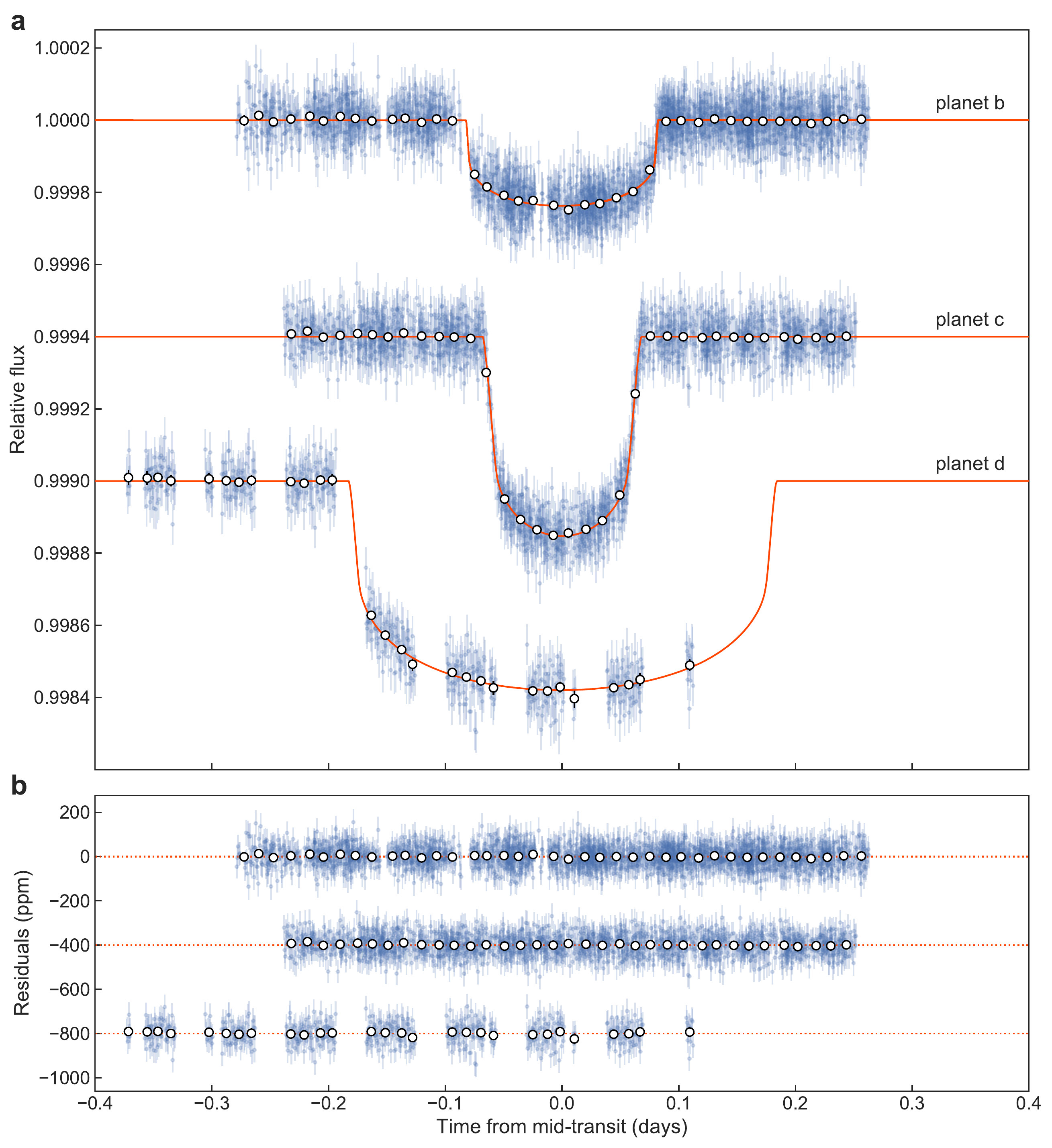}
\caption{\begin{small}\textbf{Figure 1.} \textbf{\textit{CHEOPS} transit photometry of the $\nu^2$ Lupi planets.} \textbf{a}, Corrected and phase-folded \textit{CHEOPS} transit photometry of $\nu^2$ Lupi b (\textit{top}), \hbox{c (\textit{middle}),} and d (\textit{bottom}). The blue dots show the unbinned measurements, with error bars corresponding to the quadratic sum of the formal photometric errors and the fitted extra jitter term. Open circles show the light curves binned into 20-minute intervals. The best-fit transit models from our global analysis are shown as orange lines. \textbf{b}, Corresponding residuals. In both panels, the light curves corresponding to planets c and d are shifted vertically for clarity. \end{small} \label{CHEOPS_LCs_comb}}
\end{figure*}

\begin{figure*}
\centerline{\includegraphics[width=1.0\hsize]{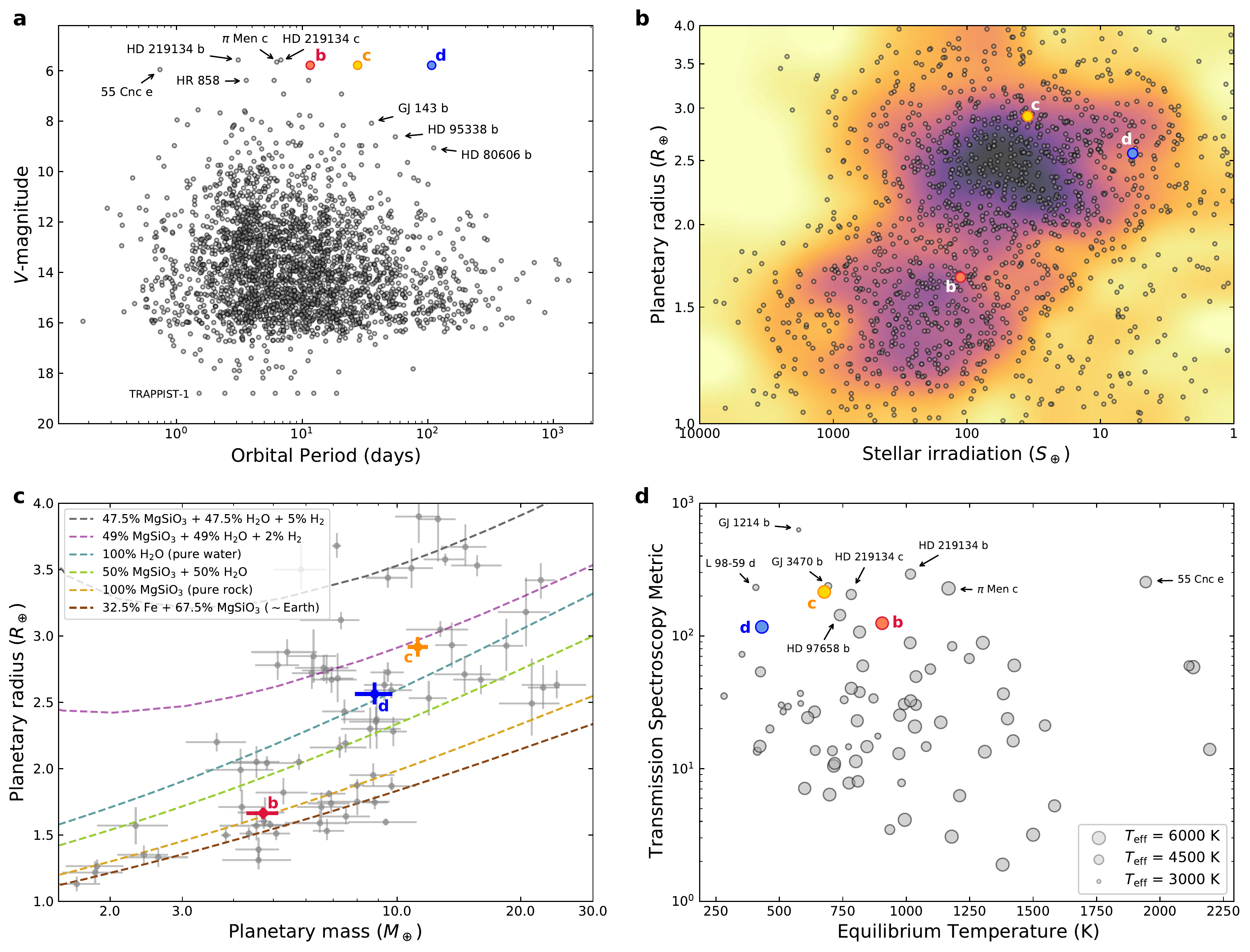}}
\caption{\begin{small}\textbf{Figure 2.} \textbf{The $\nu^2$ Lupi planets in the context of other known transiting exoplanets.} Data on known exoplanets were taken from NASA Exoplanet Archive (\url{https://exoplanetarchive.ipac.caltech.edu/}) on 18 February 2021. We only considered planets with radius uncertainties less than 20$\%$ here. \textbf{a}, $V$-magnitude versus orbital period. \textbf{b}, Planetary radius versus stellar irradiation for planets with radii below 4 $R_{\oplus}$. The background colors describe the density of data points, from yellow for empty regions of the diagram to violet for densely-populated ones. \textbf{c}, Mass-radius diagram for the planets in this sample that also have mass uncertainties less than 20$\%$. The dashed lines show theoretical mass-radius curves for some idealized compositions\cite{2019PNAS..116.9723Z}. \textbf{d}, For the same subset of planets, the transmission spectroscopy metric of \hbox{ref. \cite{2018PASP..130k4401K}} is shown as a function of equilibrium temperature (computed assuming an efficient heat redistribution and a null Bond Albedo). The size of the symbols is proportional to the host star effective temperature.\end{small} \label{fig:diagrams}}
\end{figure*}

Since transit and RV data only yield the parameters of the planets relative to those of their star, we performed a detailed characterization of the host star (see Methods), whose main derived parameters are given in Table~\ref{table_glob}. We carried out our global data analysis using the \texttt{juliet} package\cite{2019MNRAS.490.2262E}, which allows the joint modeling of transits and radial velocities simultaneously with signals of non-planetary origin (instrumental systematics, stellar variability) to ensure a full propagation of the uncertainties on the derived system parameters. A key feature of \texttt{juliet} is that it uses nested sampling algorithms\cite{dynesty} to explore the parameter space, which also allows one to perform model comparison via Bayesian evidences. We first analysed each dataset individually, to explore for each of them a large range of correlated noise models and select the best one, based on Bayesian evidence (see Methods). During this process, we detected in the \textit{CHEOPS} light curves, besides typical instrumental noise phased with the spacecraft roll angle, some extra higher-frequency correlated noise which we attribute to stellar granulation and oscillations. We modeled this stellar noise using a Gaussian Process (see Methods) and studied it in detail in an independent analysis (see below). We then performed several global analyses assuming circular or eccentric orbits for some or all the planets (see Methods). We found the simplest model with circular orbits for the three planets to be the one with the highest Bayesian evidence and thus adopted it as our nominal solution. We also explored the data for transit timing variations but did not detect any (see Methods). Table~\ref{table_glob} presents the results of our global analysis, which significantly refines the system parameters (see Methods). The best-fit models for the individual light curves are shown in Supplementary Figures 1 (\textit{CHEOPS}) and 4 (\textit{TESS}). The corrected light curves, phase-folded for each planet, are presented together with the best-fit transit models in Figure~\ref{CHEOPS_LCs_comb} for \textit{CHEOPS} and Supplementary Figure 5 for \textit{TESS}. Finally, the phase-folded RVs are shown in Supplementary Figure 6, together with the best-fit RV model for each planet. To test our results, we explored the orbital stability of the system for masses and eccentricities around our derived solution and found it to be very stable (see Methods).

As mentioned above, the exquisite precision of our \textit{CHEOPS} photometry ($\sim$15 ppm with a 10-minute binning) makes it sensitive to stellar granulation and oscillations. To characterize this stellar signal, we analysed the power spectral density (PSD) of the photometric residuals obtained after subtracting our best-fit transit and instrumental models (see Methods and Supplementary Figure 10). We measured for the stellar oscillations a frequency at maximum power \hbox{$\nu_{max}$ = \hbox{$2710 \pm 77\:\mu$Hz}} and a flicker index $\alpha_g = 1.14 \pm 0.22$ (slope of the PSD associated with granulation) that are in agreement with the values expected from our derived stellar parameters and with the trends observed for \textit{Kepler} stars\cite{2020A&A...636A..70S} (Supplementary Figure 11). This asteroseismic detection is in good agreement with, and even exceeds, the expectations from ref. \cite{2018A&A...620A.203M}, thus spectacularly demonstrating the potential of \textit{CHEOPS} for asteroseismology.

With a radius of $1.664\pm0.043\:R_{\oplus}$ and a stellar irradiation of $111.6_{-6.8}^{+7.3}\:S_{\oplus}$, $\nu^2$ Lupi b lies near the inner edge of the well-known radius valley\cite{2017AJ....154..109F}, while planets c ($2.916_{-0.073}^{+0.075}\:R_{\oplus}$, $35.1_{-2.1}^{+2.3}\:S_{\oplus}$) and d ($2.562_{-0.079}^{+0.088}\:R_{\oplus}$, $5.74_{-0.35}^{+0.38}\:S_{\oplus}$) are located on the other side (Figure~\ref{fig:diagrams}b). This gap in the planetary radius distribution separates predominantly rocky planets from larger volatile-rich (hydrogen, helium, water) sub-Neptunes. The $\nu^2$ Lupi planets seem to fit well this picture: planet b has an Earth-like bulk density of $1.02_{-0.12}^{+0.13}\:\rho_{\oplus}$, while planets c and d have significantly lower densities of $0.453_{-0.041}^{+0.045}\:\rho_{\oplus}$ and $0.522_{-0.072}^{+0.078}\:\rho_{\oplus}$, respectively, implying that they contain water and/or gas (Figure~\ref{fig:diagrams}c). Several theories have been put forward to explain the radius valley, such as photo-evaporation\cite{2017ApJ...847...29O,2018MNRAS.479.4786V}, core-powered mass loss\cite{2018MNRAS.476..759G}, or combined formation and evolution effects\cite{2020A&A...643L...1V}. With three transiting planets spanning the valley, the $\nu^2$ Lupi system provides a valuable opportunity to test these scenarios.

\begin{figure*}
\centerline{\includegraphics[width=0.9\hsize]{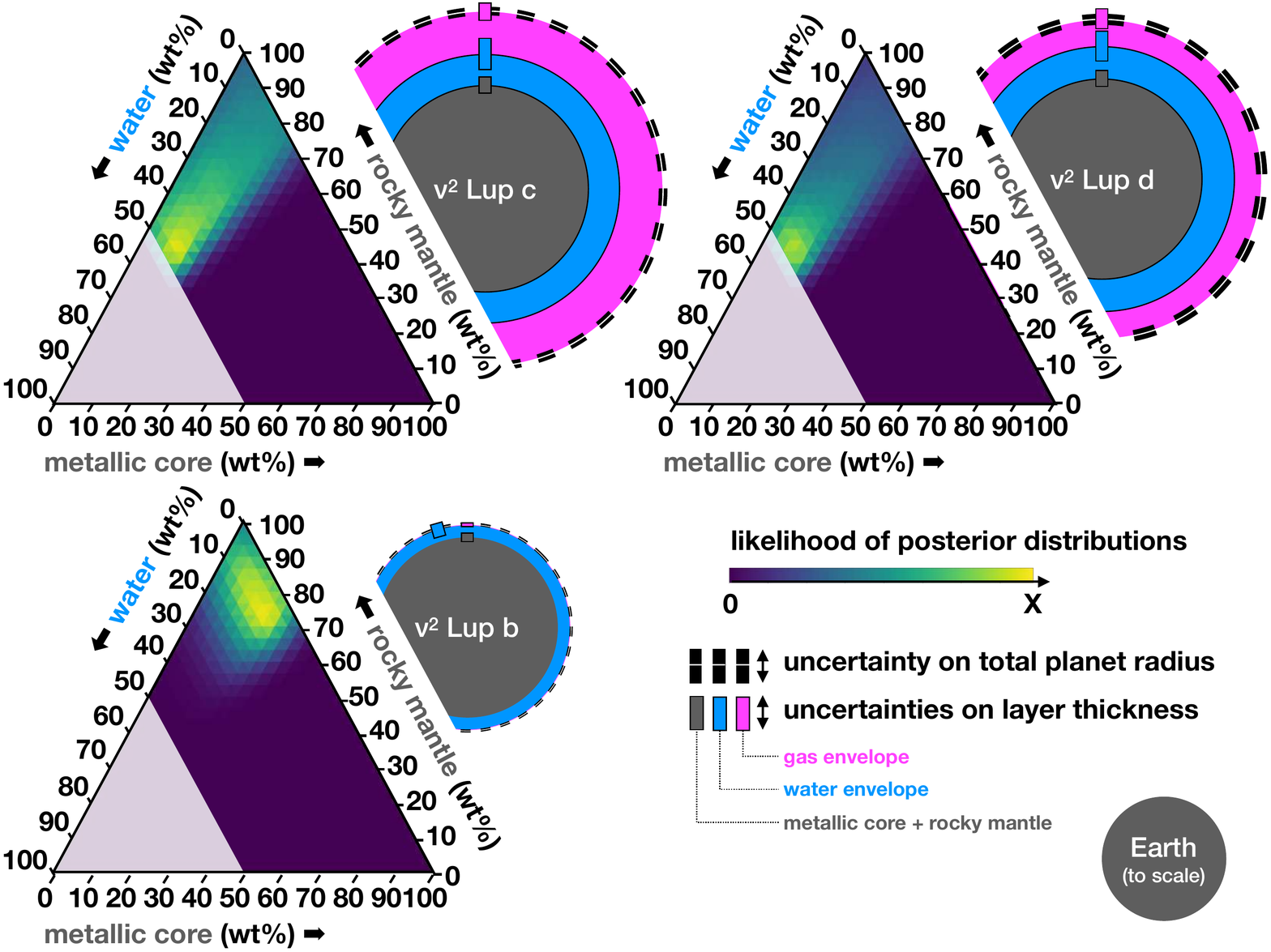}}
\caption{\begin{small}\textbf{Figure 3.} \textbf{Internal structures of the $\nu^2$ Lupi planets.}
Present-day internal structures returned by our Bayesian analysis\cite{Dorn15,Dorn17} (see Methods), assuming four layers (iron-sulfur core, silicate mantle, water layer, H/He envelope). For each planet, a ternary diagram shows the mass fractions of the iron core, silicate mantle, and water layer. The yellow (resp. dark violet) colors represent zones of highest (resp. lowest) posterior probability distribution, and the grey area (lower left part) represents the zone that is excluded by the assumed prior (water mass fraction smaller than 50\% so that it cannot be larger than the icy fraction inside planetary solid building blocks, see Methods and references therein). Next to each ternary diagram, we also show an illustration representing the radius fractions of the core+mantle (dark grey), water layer (blue), and gas envelope (magenta), corresponding to the medians of the posterior distributions.
\end{small} \label{triangles}}
\end{figure*}

To go one step further, we performed a detailed Bayesian analysis (see Methods) to derive the joint posterior distribution of the present-day internal structures of the three planets\cite{Dorn15,Dorn17} (\hbox{Figure \ref{triangles}} and Supplementary Figures 12 to 14), assuming four layers (iron-sulfur core, silicate mantle, water layer, gas envelope). The innermost planet is found to be mostly dry (water mass equal to $0.57^{+0.60}_{-0.49}\:M_{\oplus}$, i.e. $12.6_{-11.0}^{+14.7}$ weight percent wt\%) and gas-poor (gas mass less than $10^{-5} M_\oplus$), whereas the two outer planets have a similarly massive water layer ($2.81^{+2.52}_{-2.52}\:M_{\oplus}$, i.e. \hbox{$25.01_{-22.4}^{+22.2}$ wt\%} for \hbox{planet c}; $2.31^{+2.09}_{-2.13}\:M_{\oplus}$, i.e. $26.8_{-23.7}^{+21.2}$ wt\% for \hbox{planet d}) and a small but non-negligible amount of gas ($0.13^{+0.103}_{-0.078}\:M_{\oplus}$, i.e. $1.2_{-0.70}^{+0.91}$ wt\% for planet c; $0.058^{+0.069}_{-0.050}\:M_{\oplus}$, i.e. $0.66_{-0.58}^{+0.79}$ wt\% for planet d). The dichotomy between the derived amounts of water is consistent with a formation by migration, where the innermost planet would start its formation inside the ice line (located a few au from the star for typical protoplanetary disks) whereas the two outer planets would have spent at least part of their formation time outside the ice line, therefore accreting icy bodies. 

With regard to the derived amounts of gas, planetary atmospheric evolution calculations\cite{kubyshkina2019a,kubyshkina2019b} indicate that the innermost planet b was subject to significant atmospheric loss, while planets c and d did not suffer strong evaporation (see Methods and Supplementary Figure 15). The two outer planets are indeed sufficiently massive and far away from the host star to be only little affected by mass loss throughout their evolution. Therefore, the current low gas content returned by our internal structure modelling for these two planets is likely of primordial origin. In the standard core-accretion model, planets start to accrete substantial amounts of gas when they reach the critical mass, which is of the order of $\sim$10 $M_{\oplus}$ but also depends strongly on different parameters, in particular the accretion rate of solids (higher accretion rates translating into larger critical masses)\cite{1999ApJ...521..823P}. The structure of the two outer planets, as observed today, being likely primordial, these two objects provide a very important anchor for planet formation models, as they indicate that neither of them reached the critical mass during their formation. These two planets, by giving access to both the core mass and gas-to-core ratio for two objects \textit{in the same system} will provide strong constraints on the understanding of the formation of sub-critical planets. Since the presence of large gas envelopes hinders habitability\cite{2014A&A...561A..41A}, the $\nu^2$ Lupi system, with its two sub-critical outer planets, also provides an interesting case study for numerical models targeting the emergence of habitable worlds.

A thorough characterization of the system will require atmospheric measurements with, for example, the upcoming \textit{James Webb Space Telescope} or future ground-based extremely large telescopes. Based on the Transmission Spectroscopy Metric (TSM) of ref. \cite{2018PASP..130k4401K}, the three planets are attractive targets, with TSM values of 125, 214, and 117 for planets b, c, and d, respectively (see Methods and Figure \ref{fig:diagrams}d). In particular, $\nu^2$ Lupi d is the best target found so far around a Sun-like star for atmospheric studies in the low temperature regime (\hbox{$<$ 500 K}). It is also a potentially promising object to search for moons or rings (see Methods). Our transit detection of this exciting planet with \textit{CHEOPS} thus further increases the importance of $\nu^2$ Lupi as a cornerstone system for comparative exoplanetology studies of small worlds.



\begin{addendum}

\item[Acknowledgments] 
\textit{CHEOPS} is an ESA mission in partnership with Switzerland with important contributions to the payload and the ground segment
from Austria, Belgium, France, Germany, Hungary, Italy, Portugal, Spain, Sweden, and the United Kingdom. The Swiss participation to \textit{CHEOPS} has been supported by the Swiss Space Office (SSO) in the framework of the Prodex Programme and the Activit\'es Nationales Compl\'ementaires (ANC),
the Universities of Bern and Geneva as well as of the NCCR PlanetS and the Swiss National Science Foundation. The MOC activities have been supported by the ESA contract No. 4000124370. S.C. acknowledges the financial support by LabEx UnivEarthS (ANR-10-LABX-0023 and ANR-18-IDEX-0001). This work was supported by FCT - Funda\c{c}{\~a}o para a Ci\^encia e a Tecnologia through national funds and by FEDER (Fundo Europeu de Desenvolvimento Regional) through COMPETE2020 - Programa Operacional Competitividade e Internacionaliza\c{c}{\~a}o by these grants: UID/FIS/04434/2019; UIDB/04434/2020; UIDP/04434/2020; PTDC/FIS-AST/32113/2017 \& POCI-01-0145-FEDER- 032113; PTDC/FIS-AST/28953/2017 \& POCI-01- 0145-FEDER-028953; PTDC/FIS-AST/ 28987/2017 \& POCI-01-0145-FEDER-028987. S.C.C.B., S.G.S., and V.A. acknowledge support from FCT through FCT contracts nr. IF/01312/2014/CP1215/CT0004 and CEECIND/00826/2018 and POPH/FSE (EC) and IF/00650/2015/CP1273/CT0001. O.D.S.D. is supported in the form of work contract (DL 57/2016/ CP1364/CT0004) funded by national funds through FCT. M.J.H. acknowledges the support of the Swiss National Fund under grant 200020\_172746. A.D. and D.E. acknowledge support from the European Research Council (ERC) under the European Union's Horizon 2020 research and innovation programme (project {\sc Four Aces}; grant agreement No 724427). S.H. acknowledges CNES funding through the grant 837319.
The Spanish scientific participation in \textit{CHEOPS} has been supported by the Spanish Ministry of Science and Innovation and the European Regional Development Fund through grants ESP2016-80435-C2-1-R, ESP2016-80435-C2-2-R, ESP2017-87676-C5-1-R, PGC2018-098153-B-C31, PGC2018-098153-B-C33, and MDM-2017-0737 Unidad de Excelencia Mar\'{\i}a de Maeztu--Centro de Astrobiolog\'{\i}a (INTA-CSIC), as well as by the Generalitat de Catalunya/CERCA programme. The Belgian participation in \textit{CHEOPS} has been supported by the Belgian Federal Science Policy Office (BELSPO) in the framework of the PRODEX Program of the European Space Agency (ESA) under contract number PEA 4000131343, and by the University of Liège through an ARC grant for Concerted Research Actions financed by the Wallonia-Brussels Federation. L.D. is an F.R.S.-FNRS Postdoctoral Researcher. M.Gi. is an F.R.S.-FNRS Senior Research Associate. V.V.G. is an F.R.S.-FNRS Research Associate. M.L. acknowledges support from the Austrian Research Promotion Agency (FFG) under project 859724 “GRAPPA”. B.-O.D. acknowledges support from the Swiss National Science Foundation (PP00P2-190080). S.Sa. has received funding from the European Research Council (ERC) under the European Union's Horizon 2020 research and innovation programme (grant agreement No 833925, project STAREX). G.M.S. acknowledges funding from the Hungarian National Research, Development and Innovation Office (NKFIH) grant GINOP-2.3.2-15-2016-00003 and K-119517. For Italy, \textit{CHEOPS} activities have been supported by the Italian Space Agency, under the programs: ASI-INAF n. 2013-016-R.0 and ASI-INAF n. 2019-29-HH.0. L.B., G.P., I.P., G.S., and V.N. acknowledge the funding support from Italian Space Agency (ASI) regulated by ``Accordo ASI-INAF n. 2013-016-R.0 del 9 luglio 2013 e integrazione del 9 luglio 2015''. A.C.C. and T.G.W. acknowledge support from STFC consolidated grant number ST/M001296/1. D.G., X.B., S.C., M.F., and J.L. acknowledge their roles as ESA-appointed \textit{CHEOPS} science team members. We thank S. R. Kane for sharing some RV data before their publication and L. D. Nielsen for helping to plan the \textit{CHEOPS} observations based on her analysis of the \textit{TESS} data. We also thank M. Cretignier for his independent analysis of the HARPS RV data. 

\item[Author contributions] L.D. led the data analysis, with support from L.B., M.J.H., S.H., A.Br., A.D., P.G., N.H., M.O., and T.G.W.\:\:\: L.D. also coordinated the interpretation of the results and writing of the manuscript. D.E. designed and coordinated, with support from A.D., the \textit{CHEOPS} Early Science programme, within which these observations took place. Y.A. led the analysis of the internal structures, with support from J.H. A.Bo. and L.F. performed the atmospheric evolution simulations. L.B. carried out the TTV simulations. F.J.P. studied the orbital stability and tidal interactions. S.Sa., V.A., A.Bo., S.G.S., V.V.G., and T.G.W. performed the stellar characterization. S.Su. analysed the stellar granulation and oscillations. V.B. assessed the potential of the system for atmospheric characterization. S.C. evaluated the possibility of $\nu^2$ Lupi d having moons or rings. The other co-authors provided key contributions to the development of the \textit{CHEOPS} mission. All co-authors read and commented the manuscript, and helped with its revision. 

\item[Competing Interests] The authors declare no competing interests.

\item[Correspondence] Correspondence and requests for materials should be addressed to L.D. (ldelrez@uliege.be).

\end{addendum}


\pagebreak

\pagebreak

\begin{table*}
\centering
\caption{\textbf{Table 1.} Properties of the $\nu^2$ Lupi planetary system.}
\begin{footnotesize}
\begin{tabular}{lccc}
\hline
\textbf{Parameters} & \multicolumn{3}{c}{\textbf{Values}} \\
\hline
\vspace{0.12cm}
\bf{Star} & \multicolumn{3}{c}{\textbf{$\nu^2$ Lupi}} \\
\vspace{0.12cm}
Effective temperature, $T_{\mathrm{eff}}$ (K) & \multicolumn{3}{c}{$5664 \pm 61$} \\
\vspace{0.12cm}
Log surface gravity, log $g_{\star}$ (cgs) & \multicolumn{3}{c}{$4.39 \pm 0.11$} \\
\vspace{0.12cm}
Microturbulence, $\xi_{\mathrm{t}}$ (km/s) & \multicolumn{3}{c}{$0.85 \pm 0.02$} \\
\vspace{0.12cm}
Metallicity, $[\mathrm{M/H}]$ (dex) & \multicolumn{3}{c}{$-0.24 \pm 0.05 $} \\
\vspace{0.12cm}
Radius, $R_{\star}$ ($R_{\odot}$) & \multicolumn{3}{c}{$1.058 \pm 0.019$} \\
\vspace{0.12cm}
Mass, $M_{\star}$ ($M_{\odot}$) & \multicolumn{3}{c}{$0.87 \pm 0.04$} \\
\vspace{0.12cm}
Density, $\rho_{\star}$ ($\rho_{\odot}$) & \multicolumn{3}{c}{$0.734 \pm 0.053$} \\
\vspace{0.12cm}
Age (Gyr) & \multicolumn{3}{c}{$12.3^{+1.2}_{-2.9}$} \\
\vspace{0.12cm}
Rotation period,$^{a}$ $P_{\mathrm{rot}}$ (days) & \multicolumn{3}{c}{$23.8 \pm 3.1$} \\
\vspace{0.12cm}
Luminosity, $L_{\star}$ ($L_{\odot}$) & \multicolumn{3}{c}{${1.038 \pm 0.059}$} \\
\hline
\vspace{0.12cm}
\bf{Planets} & \bf{b} & \bf{c} & \bf{d} \\
\vspace{0.12cm}
Orbital period, $P$ (days) & 
$11.57797_{-0.00013}^{+0.00008}$ &
$27.59221 \pm 0.00011$ &
$107.245 \pm 0.050$ \\
\vspace{0.12cm}
Mid-transit time, $T_{0}$ ($\mathrm{BJD_{TDB}}-2,450,000$) & 
$8944.3726_{-0.0017}^{+0.0015}$ &
$8954.40990_{-0.00054}^{+0.00052}$ &
$9009.7759_{-0.0096}^{+0.0101}$ \\
\vspace{0.12cm}
Planet-to-star radius ratio, $R_{\mathrm{p}}/R_{\star}$ & 
$0.01442_{-0.00028}^{+0.00027}$ & 
$0.02526_{-0.00044}^{+0.00047}$ & 
$0.02219_{-0.00057}^{+0.00067}$ \\
\vspace{0.12cm}
Transit depth, $dF$ (ppm) & 
$208 \pm 8$ & 
$638_{-22}^{+24}$ & 
$492_{-25}^{+30}$ \\
\vspace{0.12cm}
Transit impact parameter, $b$ ($R_{\star}$) & 
$0.52_{-0.05}^{+0.04}$ &
$0.872 \pm 0.007$ &  
$0.41_{-0.21}^{+0.14}$ \\
\vspace{0.12cm}
Transit duration, $W$ (hours) & 
$3.935_{-0.058}^{+0.093}$ & 
$3.251_{-0.031}^{+0.033}$ & 
$8.87_{-0.63}^{+0.56}$ \\
\vspace{0.12cm}
Orbital inclination, $i$ (degree) & 
$88.49_{-0.15}^{+0.17}$ & 
$88.571_{-0.045}^{+0.042}$ & 
$89.73_{-0.09}^{+0.14}$ \\
\vspace{0.12cm}
Orbital eccentricity, $e$ & 
0 (fixed, $<$0.17$^{b}$) & 
0 (fixed, $<$0.08$^{b}$) & 
0 (fixed, $<$0.25$^{b}$) \\
\vspace{0.12cm}
RV semi-amplitude, $K$ (m/s) & 
$1.46 \pm 0.12$ & 
$2.61 \pm 0.12$ & 
$1.30 \pm 0.13$ \\
\vspace{0.12cm}
Orbital semi-major axis, $a$ (au) &  
$0.0964 \pm 0.0028$ & 
$0.1721 \pm 0.0050$ & 
$0.425 \pm 0.012$ \\
\vspace{0.12cm}
Scale parameter, $a/R_{\star}$ & 
$19.60_{-0.46}^{+0.45}$ & 
$34.97_{-0.82}^{+0.80}$ & 
$86.46_{-2.02}^{+1.96}$ \\
\vspace{0.12cm}
Stellar irradiation, $S_{\mathrm{p}}$ ($S_{\oplus}$) & 
$111.6_{-6.8}^{+7.3}$ & 
$35.1_{-2.1}^{+2.3}$ & 
$5.74_{-0.35}^{+0.38}$ \\
\vspace{0.12cm}
Equilibrium temperature,$^{c}$ $T_{\mathrm{eq}}$ (K) & 
905 $\pm$ 14 & 
$677 \pm 11$ & 
431 $\pm$ 7 \\
\vspace{0.12cm}
Radius, $R_{\mathrm{p}}$ ($R_{\oplus}$) & 
1.664 $\pm$ 0.043 & 
$2.916_{-0.073}^{+0.075}$ & 
$2.562_{-0.079}^{+0.088}$ \\
\vspace{0.12cm}
Mass, $M_{\mathrm{p}}$ ($M_{\oplus}$) & 
$4.72 \pm 0.42$ & 
$11.24_{-0.63}^{+0.65}$ & 
$8.82_{-0.92}^{+0.93}$ \\
\vspace{0.12cm}
Mean density, $\rho_{\mathrm{p}}$ ($\rho_{\oplus}$) & 
$1.02_{-0.12}^{+0.13}$ & 
$0.453_{-0.041}^{+0.045}$ & 
$0.522_{-0.072}^{+0.078}$ \\
\hline
\end{tabular}
\end{footnotesize}
\textbf{Notes.} $ ^{a}$ From ref. \cite{2019A&A...622A..37U}. $ ^{b}$ 2-$\sigma$ upper limits derived from our global analysis allowing all orbits to be eccentric. $ ^{c}$ $T_{\mathrm{eq}}= T_{\mathrm{eff}} \sqrt{R_\star / a} \,\, (f(1-A_{\rm B}))^{1/4}$, assuming an efficient heat redistribution ($f=1/4$) and a null Bond albedo ($A_{\rm B}=0$).
\label{table_glob}
\end{table*}


\clearpage
\begin{methods}

\subsection{\textit{CHEOPS} observations and data reduction} 
\textcolor{blue}{ }

\noindent
Six observation runs (visits) were obtained with \textit{CHEOPS} between 4 April and 6 July 2020. The log of these observations is presented in Supplementary Table 1. Each visit lasted between 11 and 12.8 hours, so as to cover transits of planets b and c (duration of $\sim$3.9 and $\sim$3.2 hours, respectively) together with substantial out-of-transit baseline. Due to \textit{CHEOPS} low Earth orbit (altitude \hbox{$\sim$ 700 km}), the data show some gaps corresponding to Earth occultations or passages through the South Atlantic Anomaly, resulting in observing efficiencies between 49 and 60\%. Due to the target's brightness ($V$=5.78), we used a short exposure time of 1.7 s and co-added on-board 26 exposures, yielding a cadence of 44.2 s. For a detailed description of \textit{CHEOPS} instrumentation, technicalities of its observations, and on-board processing, see \hbox{ref. \cite{willy}}.

\noindent
The data were automatically processed with the \textit{CHEOPS} data reduction pipeline\cite{2020A&A...635A..24H} (DRP, version 12), for which a detailed description can be found in \hbox{ref. \cite{2020A&A...635A..24H}}. In short, the DRP calibrates the raw images (event flagging, bias and gain corrections, linearization, dark current and flat field corrections), corrects them for environmental effects (smearing trails, depointing, background, and stray light), and performs aperture photometry to extract target fluxes for various apertures. For all the visits, we found a minimal light curve RMS with the default aperture of 25 pixels. The resulting light curves are shown in Supplementary Figure 1 (upper panels for each visit). Owing to the extended and irregular shape of the \textit{CHEOPS} Point Spread Function\cite{willy} (PSF) and the fact that the field rotates around the target along the spacecraft's orbit, nearby background stars can introduce a time-variable flux contamination in the photometric aperture, in phase with the spacecraft roll angle. The DRP also estimates this contamination by using \textit{Gaia} DR2 catalog\cite{GaiaCollaboration2018} to simulate \textit{CHEOPS} images of the field of view. For our $\nu^2$ Lupi observations, this contamination was very small, varying between 0.025 and 0.030\% of the target's flux.  

\noindent
In the light curve of the fifth visit (8-9 June 2020), we serendipitously detected a $\sim$500 ppm transit-like flux drop which started during the targeted transit of planet c, and lasted for the rest of the visit (see Supplementary Figure 1, bottom left panel). We carefully checked the data for systematics and found this signal to be very robust. As shown in Supplementary Figure 2, it does not show any correlation with the size of the photometric aperture, nor with any instrumental or environmental parameter (background, position of the target's PSF centroid on the CCD, various voltages and temperatures, dark current, etc). Neither can the signal be ascribed to cosmic rays or telegraphic pixels (noisy unstable pixels whose state randomly flips between a normal behavior and an arbitrary high response). Having ruled out any systematics as the origin of this signal, we then turned towards another possible culprit, \hbox{$\nu^2$ Lupi d}. This planet is the third one detected in HARPS radial velocities (RVs) by \hbox{ref. \cite{2019A&A...622A..37U}}, who reported an orbital period of $\sim$107.6 d and a minimum mass ($M_{\mathrm{p}}\:\mathrm{sin}\:i$, where $M_{\mathrm{p}}$ is the mass of the planet and $i$ is the orbital inclination) of $\sim$8.6 $M_{\oplus}$. Ref. \cite{2020AJ....160..129K} did not find any evidence for a transit of planet d in the \textit{TESS} data but also noted that this was totally expected, since the \hbox{28-day} long observations actually did not cover any inferior conjunction of the planet, as predicted from the RVs. Based on the orbital solution of \hbox{ref. \cite{2019A&A...622A..37U}}, the inferior conjunction which is the nearest in time to the fifth \textit{CHEOPS} visit was predicted at 2,459,023.21 $\pm$ 10.33 BJD. The transit-like signal that we detected started around 2,459,009.59 BJD, thus at 1.3$\sigma$. We demonstrate below (see Section ``On the origin of the \textit{CHEOPS} single transit event'') that it most likely originates from a transit of \hbox{$\nu^2$ Lupi d}.

\subsection{\textit{TESS} observations and data reduction} 
\textcolor{blue}{ }

\noindent
$\nu^2$ Lupi was observed by \textit{TESS}\cite{2015JATIS...1a4003R} during Sector 12 of its primary mission, from 21 May until 18 June 2019. During these 28 days, \textit{TESS} saved and downlinked images of $\nu^2$ Lupi every two minutes, resulting in a total of 20,119 photometric measurements. Ref. \cite{2020AJ....160..129K} recently reported the detection of two transits of planet b and one of planet c in these data. As mentioned above, these observations did not cover any transit of planet d.

\noindent
Following a similar approach to the one used by ref. \cite{2020AJ....160..129K}, we extracted our own custom light curve from the \textit{TESS} pixel files, in order to conduct a careful treatment of spacecraft systematics. Using the {\tt lightkurve} python package\cite{Lightkurve2018}, we retrieved the calibrated 2-minute target pixel image files from the Mikulski Archive for Space Telescopes (\url{https://archive.stsci.edu}) using the default quality bitmask. We extracted light curves for twenty different apertures centered on the target, which were then background-corrected by subtracting the sky contribution determined using a custom background mask (Supplementary Figure 3). We chose the photometric aperture minimizing the 1-hour Combined Differential Photometric Precision (CDPP) metric\cite{2012PASP..124.1279C}, in this case a circular aperture with a radius of 4.6 pixels. We corrected the extracted fluxes for the contamination from other faint sources in the aperture, based on their \textit{TESS} magnitudes from the \textit{TESS} Input Catalog\cite{2019AJ....158..138S}. We note that despite the large pixel scale of \textit{TESS} (21 arcseconds/pixel), $\nu^2$ Lupi is so much brighter than other nearby stars that the contamination ($\rm{Flux_{contaminants}/Flux_{target}}$) is only 0.95\% in this case.
 
\noindent
As can be seen in the upper panel of Supplementary Figure 4, the resulting light curve suffers from instrumental systematics, mostly related to the pointing jitter of the spacecraft. To correct for these systematics, we retrieved the engineering quaternion measurements for Camera 1 (which observed \hbox{$\nu^2$ Lupi}), which are two-second-cadence vector time series that describe the spacecraft attitude based on observations of a set of guide stars (\url{https://archive.stsci.edu/missions/tess/engineering/}). For each of the three vector components, we computed the means and standard deviations of the measurements within each 2-minute science image. We also retrieved the various cotrending basis vectors (CBVs, \url{https://archive.stsci.edu/tess/bulk_downloads/bulk_downloads_cbv.html}) computed by the Pre-Search Data Conditioning module\cite{2012PASP..124.1000S,2014PASP..126..100S} of the \textit{TESS} Science Processing Operations Center pipeline\cite{2016SPIE.9913E..3EJ}. We then decorrelated our light curve against the quaternion and CBV time series using Bayesian linear regression, while masking the in-transit points in the fit. We tested many different combinations of regressors and kept the one minimizing the Bayesian Information Criterion\cite{1978AnSta...6..461S}, which notably comprised the first and second-order quaternion time series as well as the high-frequency (band 3) CBVs. At the end of the process, one thousand samples were drawn from the posterior distribution of the regression coefficients to estimate the uncertainty of the model, which was then added quadratically to the error bars of the corrected data points. The resulting decorrelated light curve is shown in the second panel of Supplementary Figure 4. The 1-hour CDPP improved from 62 to 37 ppm/hour, which is comparable with the precision achieved by ref. \cite{2019ApJ...881L..19V} for HR 858, a similarly bright star. We used this decorrelated light curve in our subsequent global analysis.

\subsection{Archival radial velocities} 
\textcolor{blue}{ }

\noindent
In our global analysis, we also included 246 previously published\cite{2019A&A...622A..37U,2020AJ....160..129K} RV measurements that were obtained between 27 May 2004 and 4 August 2017 with the HARPS spectrograph on ESO 3.6-m telescope (La Silla, Chile). Among these, the last six measurements were acquired after an instrument upgrade and we thus treated them as an independent dataset in our analysis. These RV data are shown in the upper panel of Supplementary Figure 6. For details about the instrument, observations, and data reduction, see ref. \cite{2019A&A...622A..37U,2020AJ....160..129K} and references therein.

\noindent
Ref. \cite{2020AJ....160..129K} also published 169 RV measurements obtained with the UCLES spectrograph on the 3.9-m Anglo-Australian Telescope (Siding Spring, Australia), as well as 43 RV measurements obtained with the HIRES spectrograph on the 10-m Keck I telescope (Mauna Kea, Hawaii). However, these data have significantly larger uncertainties than the HARPS data (mean measurement uncertainties of 1.27 and 1.17 m/s for UCLES and HIRES vs 0.42 m/s for HARPS) and show a significantly larger scatter (5.7 and 4.5 m/s for UCLES and HIRES vs 2.7 m/s for HARPS - see Figure 1 of ref. \cite{2020AJ....160..129K}). After some preliminary analyses of the RV data (see below), it appeared that including the UCLES and HIRES data did not improve the fit, which is mostly dominated by the precise HARPS data. Thus, we only included the HARPS data in our final global analysis.

\subsection{Host star properties}
\textcolor{blue}{ }

\noindent
The main stellar parameters are presented in the upper part of Table~\ref{table_glob}. The spectroscopic parameters (effective temperature $T_{\mathrm{eff}}$, surface gravity $\log g_{\star}$, microturbulence $\xi_{\mathrm{t}}$, and metallicity [Fe/H]) were taken from ref. \cite{Sousa2008}. These parameters were derived with the same methodology used in ARES+MOOG, which was recently described in detail in ref. \cite{Sousa2014, Santos2013}. The uncertainties were updated following the discussion in ref. \cite{Sousa2011} about precision versus accuracy errors. In summary, the method starts with the measurement of equivalent widths of iron lines using the ARES code\cite{Sousa2007, Sousa2015}. Then a minimization process is applied to find the ionization and excitation equilibrium and converge to the best set of spectroscopic parameters. This process makes use of a grid of Kurucz model atmospheres\cite{Kurucz1993} and the radiative transfer code MOOG\cite{Sneden1973}. 

\noindent
Stellar atmospheric abundances of several refractory elements are given in Supplementary Table 2. They were computed using the same tools and models of atmospheres as for the determination of the stellar atmospheric parameters (see above). In our analysis, we followed the classical curve-of-growth method under assumption of local thermodynamic equilibrium\cite{Adibekyan2012, Adibekyan2015}. Our results show that the star is enhanced in $\alpha$-elements (Mg, Si, Ti, and Ca) relative to iron. 

\noindent
Using the stellar parameters derived above and {\em Gaia}, 2MASS, and {\em WISE} broadband photometry, we determined the radius of $\nu^2$ Lupi via the infrared flux method (IRFM)\cite{Blackwell1977} which, via the comparison of observed fluxes with synthetic photometry of atmospheric models, allows for the calculation of stellar effective temperature and angular diameter, and thus stellar radius when combined with the parallax. Following the retrieval of {\it Gaia} G, G$_{\rm BP}$, and G$_{\rm RP}$, 2MASS J, H, and K, and {\it WISE} W1 and W2 fluxes and relative uncertainties\cite{Skrutskie2006,Wright2010,GaiaCollaboration2018}, we employed a MCMC approach to compare the observed photometry with synthetic values derived by convolving stellar atmospheric models\cite{Castelli2003} with the throughput of the broadband bandpasses, taking the stellar parameters determined above as normal priors on the synthetic spectral energy distributions used. Via this method, we derived a stellar radius $R_{\star} = 1.058 \pm 0.019\:R_{\odot}$ that is in agreement (1.8$\sigma$) with the one reported in previous work\cite{2020AJ....160..129K} ($1.012 \pm 0.018\:R_\odot$), with differences potentially arising from differing methods used to derive $R_{\star}$ or the careful treatment of stellar metallicity conducted in our study.

\noindent
We derived the stellar age and mass from stellar models calibrated to reproduce the  aforementioned IRFM radius, metallicity accounting for alpha-enrichment, and effective temperature. Two sets of stellar parameters were computed with different stellar evolution codes, respectively using the CLES code\cite{Scuflaire08} and the PARSEC stellar tracks and isochrones\cite{marigo17}. In the first case, with CLES, the optimal stellar parameters were determined following a Levenberg-Marquardt local minimisation scheme. The CLES stellar models adopt the solar chemical mixture\cite{Asplund09}, OPAL opacities\cite{Iglesias96}, FreeEOS equation of state\cite{Irwin12}, and nuclear reactions from ref. \cite{Adelberger11}. In the second case, with PARSEC, output parameters were inferred using the Isochrone placement interpolation scheme\cite{bonfanti15,bonfanti16}. The PARSEC models adopt the solar-scaled composition given by ref. \cite{caffau11}, nuclear reaction rates considering the JINA REACLIB database\cite{cyburt10}, OP\cite{seaton05} and {\AE}SOPUS\cite{marigo09} opacities, and FreeEOS equation of state\cite{Irwin12}.
For each method, we assumed that the stellar mass and age probability distribution functions follow Gaussians whose medians and standard deviations correspond to the values and errors returned by the fit. We then combined the results of the two approaches by merging the respective Gaussian distributions. The medians and standard deviations of the resulting combined distributions were adopted as the final stellar mass and age, and their associated errors. We find a stellar mass of $0.87 \pm 0.04\:M_{\odot}$ and an old age of \hbox{$12.3^{+1.2}_{-2.9}$ Gyr}, consistent with the star's thick disk kinematics\cite{2007A&A...461..171E} and low iron abundance ($-0.34\pm0.04$ dex).

\subsection{Global data analysis and derivation of the system parameters}
\textcolor{blue}{ }

\noindent
To determine the system parameters, we performed a combined analysis of the \textit{CHEOPS} and \textit{TESS} transit photometry, together with the HARPS radial velocity data. For this purpose, we used the publicly available \texttt{juliet} library\cite{2019MNRAS.490.2262E} (\url{https://github.com/nespinoza/juliet}), which is built over \texttt{batman}\cite{batman} for the modeling of transits and \texttt{radvel}\cite{radvel} for radial velocities. A key feature of \texttt{juliet} is that it uses nested sampling algorithms (in this work, \texttt{dynesty}\cite{dynesty}) to explore the parameter space, which also allows one to perform model comparison via Bayesian evidences. For this analysis, we assumed that the \textit{CHEOPS} single transit event was caused by \hbox{$\nu^2$ Lupi d} (see Section ``On the origin of the \textit{CHEOPS} single transit event'' below). 

\noindent
\texttt{juliet} is very versatile and allows for a variety of parametrizations. In our analysis, we used the following model parameters:
\begin{itemize}
    \item For each planet, the orbital period $P$ and the mid-transit time $T_{0}$. 
    \item For each planet, the parameters $r_1$ and $r_2$ as introduced by ref. \cite{2018RNAAS...2..209E}. This parametrization allows to efficiently explore the physically plausible zone in the ($b, p$) plane, where $p$ is the planet-to-star radius ratio ($R_{\mathrm{p}}/R_\star$) and $b$ is the transit impact parameter (see ref. \cite{2018RNAAS...2..209E} for details).
    \item The stellar density $\rho_{\star}$ which, together with the orbital period $P$ of each planet, defines through Kepler's third law a value for the scaled semi-major axis $a/R_{\star}$ of each planet. This parametrization offers the advantage of defining a single common value of the stellar density for the system rather than fitting for the scaled semi-major axis of each planet, thus reducing the number of fitted parameters. We placed a normal prior (\hbox{$\mathcal{N}$(1035, 76$^2$) kg/m$^{3}$}) on the stellar density based on the stellar mass ($M_{\star} = 0.87 \pm 0.04\:M_{\odot}$) and radius ($R_{\star} = 1.058 \pm 0.019\:R_{\odot}$) that we previously derived.
    \item For each bandpass (\textit{CHEOPS} and \textit{TESS}), two transformed limb-darkening coefficients ($q_1$, $q_2$), referring to the formalism introduced by ref. \cite{2013MNRAS.435.2152K}. This parametrization allows an efficient uninformative sampling from the physically plausible ($u_1$, $u_2$) parameter space, where $u_1$ and $u_2$ are the quadratic limb-darkening coefficients.
    \item For each planet, $\sqrt{e}\:\mathrm{sin}\,\omega$ and $\sqrt{e}\:\mathrm{cos}\,\omega$, where $e$ is the orbital eccentricity and $\omega$ is the argument of periastron, as well as the RV semi-amplitude $K$. 
\end{itemize}
Together, these model parameters describe the planetary signals present in the photometric and RV data. To these should be added some ``nuisance'' parameters describing the instrumental or astrophysical noise. \texttt{juliet} allows indeed to model the planetary signals simultaneously with some possibly underlying correlated noise, ensuring this way a full propagation of the uncertainties on the derived system parameters. This is done using either linear models (against any relevant external parameter) or Gaussian Processes (GPs; \texttt{celerite}\cite{celerite}, \texttt{george}\cite{george}).

\noindent
We first performed individual analyses of each of our light curves, in order to select for each of them the best correlated noise model, based on Bayesian evidence. We explored a large range of models for the \textit{CHEOPS} light curves, consisting of first to fourth order polynomials in the recorded external parameters (most importantly: time, background level, position of the PSF centroid, spacecraft roll angle), as well as GPs against time, roll angle, or a combination of both. In this process, we only selected a more complicated model over a simpler one if the difference in its Bayesian log evidence ($\Delta$ ln $Z$) was greater than two\cite{2019MNRAS.490.2262E,2008ConPh..49...71T}. We found out that linear functions of sin($n\phi$) and cos($n\phi$), where $n$ = 1, 2, 3 (depending on the visit, see Supplementary Table 3) and $\phi$ is the spacecraft roll angle\cite{willy}, are favored for all visits and account for most of the instrumental noise (shown in turquoise in the top panels of Supplementary Figure 1). On top of that, we noticed in the photometric residuals some higher-frequency correlated noise which we attribute to stellar granulation and oscillations (see below for a detailed study of the power spectrum of the residuals and characterization of the stellar noise). If not properly accounted for, such a stellar noise can introduce biases in the inferred transit parameters\cite{2020A&A...636A..70S,2020A&A...634A..75B}. In our analysis, we modeled the stellar noise using a GP with a stochastically-driven damped simple harmonic oscillator (SHO) kernel\cite{celerite}, with a quality factor of $1/\sqrt{2}$. In this particular case, the SHO kernel has indeed a similar power spectral density as stellar granulation\cite{1985ESASP.235..199H,2014A&A...570A..41K}, in first approximation. Since this stellar variability is seen in all the \textit{CHEOPS} light curves, a single common SHO GP was fitted across the six \textit{CHEOPS} visits in our combined analysis (shown in green in the upper panels of Supplementary Figure 1). The \textit{TESS} light curve is not sensitive to such a short-timescale low-amplitude stellar signal, however it shows some residual systematics which we modeled using a Mat\'ern GP (shown in orange in the second panel of Supplementary Figure 4), as favored by the Bayesian evidence. As for the RVs, we simply modeled them using a sum of three Keplerians (to account for the three planets), with a different zero point for each dataset (pre-upgrade and post-upgrade). No correlated noise model was necessary in this case. Finally, we also fitted for each of our photometric and RV datasets an extra jitter term which was added quadratically to the error bars of the data points, to account for any underestimation of the uncertainties or any excess noise not captured by our modeling.

\noindent
Altogether, this yields a total of 70 free model parameters, of which 26 describe the planetary signals and the other 44 are nuisance parameters. These parameters are listed together with their priors in Supplementary Table 3. To aid convergence, we placed normal priors on the nuisance parameters, based on the posterior distributions returned by our individual analyses. This approach allowed us to properly account for the covariances between the physical and nuisance parameters, while propagating the prior information we had on the nuisance parameters from our individual analyses. All the physical parameters were sampled from wide uniform priors, except the stellar density for which we used the normal prior mentioned previously.

\noindent
We performed several global analyses: one assuming circular orbits ($e$ set to zero) for all three planets (hereafter the ``reference model''), one allowing the orbits of all three planets to be eccentric (hereafter the ``eccentric model''), and three more analyses allowing the orbit of one planet to be eccentric while assuming circular orbits for the other two. A comparison between the Bayesian evidences of these models is provided in Supplementary Table 4. The reference model is the one with the highest Bayesian evidence. The eccentric model is marginally disfavored compared to the reference model ($\Delta$ ln $Z$ = $-3.6$). The models allowing the orbit of only one of the planets to be eccentric are statistically indistinguishable from the reference model ($\Delta$ ln $Z$ between $-0.9$ and $-1.7$). However, given the reference model is the simplest model and it has the highest Bayesian evidence, it appears to be the best model given the data at hand. We thus adopted it as our nominal solution. 

\noindent
The posterior distributions of the main fitted parameters of our global analysis are shown in Supplementary Figure 7, while Table~\ref{table_glob} presents our results for the most relevant physical parameters of the system. The best-fit models for the individual light curves are shown in Supplementary \hbox{Figures 1} (\textit{CHEOPS}) and 4 (\textit{TESS}). The unbinned \textit{CHEOPS} light curves, which have a cadence of 44.2 s, have a residual RMS between 47 and 53 ppm. When binning into 10-minute and 1-hour intervals, we reach RMS values between 14 and 17 ppm, and between 5 and 7 ppm, respectively. For comparison, the residual RMS of the unbinned (2-minute cadence) \textit{TESS} light curve is 149 ppm. This RMS decreases to \hbox{80 and 30 ppm} when binning into 10-minute and 1-hour intervals, respectively. The corrected light curves, phase-folded for each planet, are shown in Figure~\ref{CHEOPS_LCs_comb} (\textit{CHEOPS}) and Supplementary Figure 5 (\textit{TESS}) together with the best-fit transit models. The phase-folded RVs are shown in Supplementary Figure 6, together with the best-fit RV model for each planet.

\subsection{On the refinement of the system parameters} 
\textcolor{blue}{ }

\noindent
To assess the improvement brought by our \textit{CHEOPS} data for planets b and c compared to the previous \textit{TESS} measurements\cite{2020AJ....160..129K}, we also analysed both datasets individually, in a homogeneous way, using the same methods and priors as above. For this exercise, we assumed eccentric orbits for the planets to allow a direct comparison with the results of ref. \cite{2020AJ....160..129K}. We first did a combined analysis of the \textit{TESS} light curve and the radial velocities. The results of this analysis are compared to those of ref. \cite{2020AJ....160..129K} in Supplementary Table 5 (second and third columns) for some key parameters. Our results are consistent, but our uncertainties are significantly larger, e.g. by factors of $\sim$1.7 and $\sim$2.1 for the planet-to-star radii ratios of planet b and c, respectively. These larger uncertainties are likely related to our modeling of the residual systematics with a GP which is fitted simultaneously with the transits, while ref. \cite{2020AJ....160..129K} performed a full detrending of the \textit{TESS} light curve previously to the transit fitting. We consider our approach to be more robust, as it accounts for the possible covariances between the nuisance and transit parameters, thus ensuring a proper propagation of the uncertainties on the derived transit parameters.

\noindent
We then performed a combined analysis of our six \textit{CHEOPS} visits together with the radial velocities. The results are reported in the fourth column of Supplementary Table 5. They are in good agreement with those of our \textit{TESS} data analysis and also provide tighter constraints on the transit parameters. The planet-to-star radii ratios are measured with relative precisions of 2.8\% and 2.1\% for planets b and c, respectively, a factor $\sim$2 more precise than the measurements returned by our analysis of the \textit{TESS} data. This significant refinement stems from both the higher photometric precision of \textit{CHEOPS} and the larger number of transits observed (four transits of planet b observed with \textit{CHEOPS} versus two with \textit{TESS}, three transits of planet c observed with \textit{CHEOPS} versus only one with \textit{TESS}).

\noindent
Of course, the best constraints are obtained when combining all the data (\textit{TESS}+\textit{CHEOPS}+RVs) together (last column of Supplementary Table 5). Thanks to the longer temporal baseline, the transit ephemerides are significantly refined, thus enabling efficient follow-up observations. The mid-transit times of planets b and c in May 2021 (middle of the next observing window) have now uncertainties of only 10.1 and 2.6 minutes, respectively. This is a major improvement when compared with the previous respective uncertainties of 88 and 106 minutes obtained when using the ephemerides of ref. \cite{2020AJ....160..129K}. The other transit parameters (e.g. planet-to-star radii ratios and impact parameters) are only sightly refined when comparing with the results of our \textit{CHEOPS} data analysis, which demonstrates that they are mostly constrained by \textit{CHEOPS}. The planetary radii derived from our global analysis are slightly larger than those reported by ref. \cite{2020AJ....160..129K}. This is due to the combined effect of both our slightly larger planet-to-star radii ratios and stellar radius (see the ``Host star properties'' section). We note however that the planetary radii are still consistent within the uncertainties.

\subsection{Transit timing variations}
\textcolor{blue}{ }

\noindent
The identification of mean-motion resonance (MMR) configurations or the detection of significant transit timing variations (TTVs) in multi-planet systems can yield valuable information about their formation and evolution\cite{Lissauer2011ApJS..197....8L}.
In particular, the TTV signal is enhanced when the planets are in, or close to,
a MMR\cite{MiraldaEscude2002ApJ...564.1019M, 
Agol2005MNRAS.359..567A, 
HolmanMurray2005Sci...307.1288H, 
Fabrycky2014ApJ...790..146F, 
Winn2015ARA&A..53..409W},
and the lower the order of the MMR, the stronger the TTV amplitude.
Looking at the period commensurability, that is the ratio of the periods of planet pairs in the system, is the first step to identify a possible MMR. In the $\nu^2$ Lupi system, both pairs of planets b-c and c-d show a period commensurability close to a third-order MMR, i.e. a 5:2 MMR for the b-c pair and a 4:1 MMR for the c-d pair.

\noindent
The common method to identify TTVs is to plot the so-called $O-C$ (Observed-Calculated) diagram, where $O$ are the observed (measured) transit times,
and $C$ are transit times computed from a linear ephemeris. To compute the expected TTV signals, we simulated the system with a N-body integrator. We used the parameters in Table~\ref{table_glob} to integrate the orbits of the three planets for 3.5 years (i.e., the nominal duration of the \textit{CHEOPS} mission)
with the \texttt{TRADES}\cite{Borsato2014, Malavolta2017a, Borsato2019} program, that allows us to compute the simulated transit times for each planet. We then extracted the semi-amplitude of the $O-C$ diagram of each planet, assuming the simulated transit times as $O$ and subtracting the transit times $C$ calculated from the linear ephemeris based on the orbital period $P$ and mid-transit time $T_{0}$ in Table~\ref{table_glob}. We found that the expected TTV semi-amplitudes are rather small: about 20~s for planets b and d, and about 40~s for planet c. 

\noindent
To explore our current dataset for possible TTVs, we ran another global analysis with \texttt{juliet}, additionally including a free TTV offset parameter for every transit of planet b (6 transits) and planet c (4 transits), while keeping their orbital period $P$ and mid-transit time $T_0$ fixed to the values given in Table~\ref{table_glob}. For the TTV offsets, we assumed uniform priors between $-15$ and $+15$ minutes. We assumed the same priors as previously (Supplementary Table 3) for the other physical and nuisance parameters, as well as circular orbits. This analysis returned uncertainties on the individual transit times of $\sim$4-5 minutes for planet b and $\sim$1-2 minutes for planet c, which did not allow us to detect any significant TTVs.

\subsection{On the origin of the \textit{CHEOPS} single transit event}
\label{4th_planet}
\textcolor{blue}{ }

\noindent
In all of the above, we assumed that the single transit event caught by \textit{CHEOPS} was caused by \hbox{$\nu^2$ Lupi d}. However, one might ask whether this transit may instead be caused by an additional yet-unknown planet in the system.

\noindent
The single transit event that we detected in the fifth \textit{CHEOPS} visit is partial (i.e. the observations did not cover the end of the transit), hence its duration is not well constrained. Additionally, the observations also did not cover the ingress which occurred during an Earth occultation. Because of that, the impact parameter and orbital period are very poorly constrained. Some more constraints on the orbital period may come from the other photometric observations, especially the 28-day long \textit{TESS} observations. A $\sim$9hr-long transit with a $\sim$500 ppm depth would have been very clearly detected in these data (see for example how the transit of $\nu^2$ Lupi c, which has a similar depth, appears clearly in Supplementary Figure 4), meaning they can be used to exclude some orbital periods. To assess the constraints brought by the photometry on the orbital period of the transiting object, we performed a combined fit of all the \textit{CHEOPS} and \textit{TESS} data with \texttt{juliet}. We assumed the same normal priors as previously for the stellar density and nuisance parameters and wide uniform priors for the planet parameters (see Supplementary Table 3). For the third transiting planet corresponding to the \textit{CHEOPS} single transit event, we assumed a uniform prior between 30 and 1000 d for the orbital period, and a uniform prior between 2,459,009.6 and 2,459,010.0 BJD for the mid-transit time. The posterior distributions of the fitted parameters for the third planet are shown in Supplementary Figure 8. A wide range of orbital periods are compatible with the photometry, with a 3-$\sigma$ upper limit of 624 d. The few blank vertical stripes seen in the corner plot for some orbital periods (e.g. around 185 d and 370 d) correspond to periods that can be excluded based on the \textit{TESS} data.

\noindent
To assess the possibility of a fourth unknown planet in the system, we also searched for some additional signals in the HARPS radial velocities. We computed the $\ell_1$ periodogram\cite{hara2017}, which searches for several periodic signals simultaneously and thus is less prone to show spurious peaks due to aliasing than a regular periodogram. The $\ell_1$ periodogram (see Supplementary Figure 9, top) 
 shows two possible peaks around 123 and 485 d, but with high false-alarm probabilities of 14.5\% and 9.6\%, respectively. We note that the signal at 123 d was already discussed by ref. \cite{2019A&A...622A..37U}, who came to the conclusion that it was more likely an artifact induced by noise in the data or interaction with the window function (period $\sim$1/3 of a year) rather than an additional planet in the system. To assess the possibility of one of these two candidates being at the origin of the \textit{CHEOPS} single transit event, we fitted a fourth planet to the HARPS RVs and checked if the posterior distributions obtained for $P$ and $T_0$ were compatible with the photometry, i.e. which percentage of the posterior samples could produce a transit in the fifth \textit{CHEOPS} visit but no transit in the \textit{TESS} data. For the 123 d signal, we found that the derived $P$ ($122.26_{-0.83}^{+0.87}$ d) and $T_0$ ($2,455,509.75 \pm 4.32$ BJD) posterior distributions are completely incompatible with the photometry: all the posterior samples that are compatible with a transit during the fifth \textit{CHEOPS} visit would have also produced a transit in the $TESS$ data. Furthermore, a dynamical stability analysis reveals that a fourth planet with a period of 123 d would make the system unstable (see next section). This candidate signal at \hbox{123 d} can thus be discarded. For the \hbox{485 d} signal, we find that 0.02\% of the derived $P$ \hbox{($481.72_{-14.88}^{+17.35}$ d)} and $T_0$ ($2,455,727.22_{-18.89}^{+20.63}$ BJD) posterior samples are compatible with the photometry. By way of comparison, this is 20 times less than the corresponding percentage of 0.4\% obtained for \hbox{$\nu^2$ Lupi d}. 

\noindent
As a complementary check, we performed two more global analyses of all the data (\textit{CHEOPS}, \textit{TESS}, HARPS) assuming a fourth planet with a period of 485 d (planet e). For one of these analyses, we assumed that planet e is responsible for the \textit{CHEOPS} single transit event (``4 planets - e transiting'' model) and that planet d was thus not detected in transit in the photometry. For the second one, we assumed the opposite, i.e. that planet d is responsible for the \textit{CHEOPS} single transit event and that planet e is not detected in the photometry (``4 planets - d transiting'' model). A comparison between the Bayesian evidences of these two models and our reference model (for which we assumed 3 planets with planet d responsible for the single transit, see above) is provided in Supplementary Table 4. The ``4 planets - e transiting'' model is marginally disfavored compared to the ``4 planets - d transiting'' model, with a $\Delta$ ln $Z$ = $-3.1$ i.e. posterior odds of $\sim$1:22 assuming equiprobable models. However, these two models are not equiprobable, planet d having also a $\sim$2.8 times higher geometric transit probability than planet e. This results in posterior odds of $\sim$1:62 in favor of the ``4 planets - d transiting'' model. This model is statistically indistinguishable from our ``3 planets - d transiting'' reference model ($\Delta$ ln $Z$ = $0.6$), which reflects the low significance of the 485 d signal in the RV data. Our simpler reference model is thus the best model given the data at hand.

\noindent
Finally, we also assessed the possibility that the \textit{CHEOPS} single transit is caused by a fourth planet that is completely undetected in the HARPS RVs. A detailed computation of RV detection limits, via injection and recovery tests, is beyond the scope of this paper. To estimate the detection limits of the HARPS dataset, we instead used the results of the radial-velocity fitting challenge reported by ref. \cite{2017A&A...598A.133D}. They showed that signals with a semi-amplitude $K \geq 7.5\: \rm{RV_{rms}}/\sqrt{N_{\rm{obs}}}$, with $\rm{N_{obs}}$ the number of data points and $\rm{RV_{rms}}$ their RMS, are typically confidently detected in RV time series. For our HARPS dataset, we have $\rm{N_{obs}}$=246 and \hbox{$\rm{RV_{rms}}$=1.37 m/s} after removing the signals of the three known planets, yielding a detection threshold $K$=0.66 m/s. From this value, we can derive a minimum planetary mass that should be detected for a given orbital period. Supplementary Figure 9 (bottom) shows this minimum planetary mass as a function of orbital period for periods up to \hbox{624 d} (3-$\sigma$ upper limit from our analysis of the photometry, see above). Since we know that the radius of the transiting object is $\sim$2.6 $R_{\oplus}$ based on its transit depth, we can estimate its mass using for example the empirical mass-radius relationship of ref. \cite{2017ApJ...834...17C}: $M_p = 1.436\:R_p^{1.70}$, thus \hbox{$\sim$7.3 $M_{\oplus}$} in this case. As shown in Supplementary Figure 9 (bottom), planets in this mass range with periods up to \hbox{$\sim$480 d} should be confidently detected in the HARPS data according to this criterion. Planets in this mass range with longer periods up to $\sim$624 d should produce low-significance peaks in the RV periodogram. This is for example the case for the 485 d candidate discussed above which has a semi-amplitude \hbox{$K \sim$0.5 m/s}, thus lower than our approximate detection threshold, but still produces a low-significance peak in the periodogram. 

\noindent
Based on all these elements, we conclude that the \textit{CHEOPS} single transit event was most likely caused by $\nu^2$ Lupi d, rather than by an additional unknown planet in the system.

\subsection{Orbital stability}
\textcolor{blue}{ }
\label{stability_section}

\noindent
In this section, we sought to test the results of our global fit by exploring the stability of the system. To achieve this, we made use of the Mean Exponential Growth factor of Nearby Orbits (MEGNO), $Y(t)$\cite{cincottasimo1999,cincottasimo2000,cincotta2003}. MEGNO is a chaos index that has been extensively used within dynamical astronomy, in particular for extrasolar planetary systems\cite{hinse2010,jenkins2019,pozuelos2020}. Its time-averaged mean value, $\langle Y(t) \rangle$, amplifies any stochastic behaviour, allowing the detection of hyperbolic regions during the integration time. Therefore, $\langle Y(t) \rangle$ allows us to distinguish between chaotic and quasi-periodic trajectories: if $\langle Y(t) \rangle \rightarrow \infty$ for $t\rightarrow \infty$ the system is chaotic; while if $\langle Y(t) \rangle \rightarrow 2$ for $t\rightarrow \infty$ the motion is quasi-periodic. We used the MEGNO implementation within the N-body integrator \texttt{REBOUND}\cite{rein2012}, which makes use of the Wisdom-Holman \texttt{WHFast} code \cite{rein2015}.

\noindent
We explored the masses and eccentricities of the planets, two parameters which have the most dramatic effect on the orbital dynamics, by building stability maps of adjacent planets. The map building entailed exploring the parameter space of $M_{b}$--$M_{c}$ and $e_{b}$--$e_{c}$ for the inner planets b and c, and $M_{c}$--$M_{d}$ and $e_{c}$--$e_{d}$ for planets c and d. Hence, we studied planetary masses ranging from 1 to 10~$M_{\oplus}$ for planet b, 5 to 15~$M_{\oplus}$ for planets c and d, and eccentricities in the range of 0.0--0.6 for all planets. In our sets of simulations, we took 10 values from each range, meaning that the size of each stability map was 10$\times$10 pixels. Moreover, for each scenario, we set 20 different random initial conditions by varying the argument of pericenter in the range of its nominal value $\pm$1$\sigma$ (based on our global eccentric analysis), and the longitude of ascending node and mean anomaly in the range 0--360~$\deg$. We then averaged these 20 initial conditions to obtain the averaged value of $\langle Y(t) \rangle$ of each pixel of the stability maps. Hence, for each stability map, we ran 2000 simulations. The integration time was set to 10$^{6}$ times the orbital period of the outermost planet d. The time step was set to 5 per cent of the orbital period of the innermost planet b. Note that we explored the nominal values of the planetary masses and eccentricities presented in Table~\ref{table_glob} up to well beyond their 5$\sigma$ ranges. This strategy allowed us to better understand the full architecture of the system, as well as its limitations. We found that the system is fully stable over the whole range of masses explored, with a $\Delta \langle Y(t) \rangle$= 2.0 $-$ $\langle Y(t) \rangle <10^{-3}$. With regard to the eccentricities, we obtained that the planets may tolerate eccentricities up to 0.3, 0.3, and 0.4 for planets b, c, and d, respectively. These masses and eccentricities are all well above their nominal values, which highlights the goodness of the solution from our global analysis.  

\noindent
As mentioned in the previous section, the RV periodogram shows two possible signals at 123 and 485 d. A fourth planet with an orbital period of 485 d would not affect the stability of the system, but a planet with a period of 123 d could make the system unstable. To test the stability of the system in such a four-planet configuration, we included a planet with an orbital period of 123 d, and then built stability maps exploring the parameter space $e_{123\rm{d}}-M_{123\rm{d}}$ (with $e_{123\rm{d}}$ and $M_{123\rm{d}}$ the eccentricity and mass of the planet at 123 d, respectively). These stability maps were built following the same procedure presented above, but we increased the number of pixels to $15\times15$. We then explored eccentricities in the range of $0.0-0.6$ and masses ranging from 1 to 15 $M_{\oplus}$. We ran two suites of simulations: (1) planets b, c, and d were given circular orbits, and (2) they were given eccentric orbits using the eccentricities derived from our global eccentric analysis. For each pixel (or equivalently, each combination of $e_{123\rm{d}}-M_{123\rm{d}}$), we ran 20 different initial conditions by randomly varying the orbital angles, and then averaged the results. That is, each stability map contained 4500 scenarios. We found that such a four-planet system would be fully unstable for times shorter than 10${^6}$ orbits of the outermost planet. Based on these results, a fourth planet at \hbox{123 d} can thus be discarded.

\subsection{Tidal interactions}
\textcolor{blue}{ }

\noindent
 The planets are close enough to their host star to experience significant tidal interactions. To investigate this aspect, we quantified the influence of tides by means of the constant time-lag model, in which a planet is considered as a weakly viscous fluid that is deformed due to gravitational effects\cite{Mignard1979TheI,Hut1981TidalSystems,Eggleton1998TheFriction}. In this context, the tidal dissipation of a given planet is defined by the product of the potential Love number of \hbox{degree 2} and the constant time lag, k$_{2}\Delta \tau$. For terrestrial exoplanets, a range of possibilities centred on Earth's dissipation factor\cite{neron1997} k$_{2}\Delta \tau_{\oplus}$=213~s, that is (0.1-10)$\times$k$_{2}\Delta \tau_{\oplus}$, is generally adopted, so as to explore a range of possible tidal behaviours\cite{demory2020,bolmont2020}. This assumption is valid for the innermost planet in the system, $\nu^2$ Lupi b, which has an Earth-like mean density (Table~\ref{table_glob}). This is not the case of $\nu^2$ Lupi c and d, whose lower densities suggest that they are likely volatile-rich and have small gaseous envelopes (see the `Internal planetary structures' Section). Hence, we assumed a tidal dissipation factor similar to Jupiter's value\cite{leconte2010}, k$_{2}\Delta \tau \sim2.5\times10^{-2}$~s, and explored a range of $1-100$ times this value\cite{nielsen2020}. We then performed a suite of simulations with the N-body code {\sc{posidonius}}\cite{Blanco-Cuaresma2017StudyingRust}, which includes the effects of tides, rotational flattening, and general relativity, using the same prescriptions given by ref. \cite{bolmont2015}. 

\noindent
We found that the innermost planet b evolved into a pseudo-rotational state rapidly; that is, into a tidally locked configuration where its orbital and rotational periods became synchronized, and its obliquity close to zero, so that its orbital axis became aligned with the spin axis of the star. This state was reached within a short time scale, of maximum $10^{4}-10^{6}$~yr. This process is slower for the two outermost planets, which should reach pseudo-rotational states after $10^{8}-10^{10}$~yr for \hbox{planet c}, and $10^{11}-10^{13}$~yr for planet d. Concerning the circularisation of the orbits, we found that it should be reached after $10^{7}-10^{9}$~yr for planet b, and after $10^{11}-10^{13}$~yr for planet c. For \hbox{planet d}, the circularisation time exceeds 10$^{14}$~yr. Dynamical tidal processes thus do not seem to be very efficient in this system. Considering that the estimated age of the system is $12.3_{-2.9}^{+1.2}$~Gyr (Table~\ref{table_glob}), we conclude that circularisation may be complete for planet b, but still ongoing for planets c \hbox{and d}. However, while eccentric orbits may not be fully discarded from our global analysis, we found the model with circular orbits for all three planets to be marginally favored. If this is the case, this would mean that planets c and d dissipated more energy than what was explored in this study, hinting that other processes such a tidal inertial waves in the convective region of the planets might be affecting the system\cite{ogilvie2004}. In this case, enhancement of the tidal dissipation rates would imply a more rapid synchronization of the planets' spins with their orbits, and faster circularisation of their orbits.

\subsection{Characterization of the stellar signal seen in \textit{CHEOPS} photometry}
\textcolor{blue}{ }

\noindent
To precisely characterize the stellar noise detected in the \textit{CHEOPS} light curves, we removed the best-fit transit and instrumental noise models from the data and analysed the power spectral density (PSD) of the resulting residuals (Supplementary Figure 10, gray curve). We clearly observe the bump of the stellar acoustic modes at high frequency, and the characteristic increase of the PSD towards the lower frequencies associated with the signature of stellar granulation\cite{1985ESASP.235..199H}.

\noindent
To characterize this stellar signal, we first performed a GP regression in the time domain based on the model described in ref. \cite{2019MNRAS.489.5764P} (dashed lines in Supplementary Figure 10). This GP model consists of a sum of three kernels chosen such as their respective PSD corresponds to a Gaussian-like envelop to describe the oscillation bump, an Harvey-like function to describe the granulation component, and a stochastic term to describe the high-frequency (photon) noise. Using an MCMC approach\cite{2013PASP..125..306F} to derive the parameters of this model, we inferred a characteristic amplitude and frequency for the granulation of $49 \pm 2$ ppm and $1026 ^{+85}_{-80}$ $\mu$Hz, respectively, and a frequency at maximum power for the stellar oscillations ($\nu_{max}$) of $2710 \pm 77$ $\mu$Hz. The latter is in agreement ($1.8\sigma$) with the $\nu_{max}$ of \hbox{$2414 ^{+141}_{-133}$ $\mu$Hz} expected from the stellar parameters given in Table~\ref{table_glob}. 

\noindent
Considering there is still some debate as to which model best describes the granulation signature (see e.g. ref. \cite{Mathur_2011, 2014A&A...570A..41K} and references therein), we tested another model based on simple power laws\cite{1999ASPC..173..297P,2020A&A...636A..70S}. 
Making the approximation that the power background follows frequency power functions, this model is defined as $\log P(\nu) = \alpha_g \log(\nu) + \beta$ with $\nu$ the frequency, $\alpha_g$ the flicker index measured between two cut-off frequencies $\nu \in [f_c, f_g]$, and $\beta$ a constant. Using again an MCMC approach\cite{2017AJ....153....3C} to derive the parameters of this power-law model, we inferred a flicker index $\alpha_g = 1.14 \pm 0.22$ between the cut-off frequencies $f_c = 1964 \pm 36$ $\mu$Hz and \hbox{$f_g = 830 \pm 130$ $\mu$Hz} (red dotted line in Supplementary Figure 10). Based on \textit{Kepler} observations, it has been shown that this flicker index and the corresponding cut-off frequencies are strongly correlated with the stellar fundamental parameters\cite{2020A&A...636A..70S}, particularly with the stellar mass, radius, and surface gravity (Supplementary Figure 11). Comparing the flicker index we inferred for $\nu^2$ Lupi (red dot with error bars) with results obtained previously for \textit{Kepler} stars (grey and black dots), we observe that it follows well the expected trends. Moreover, this first \textit{CHEOPS} measurement opens the way to studies of granulation signatures on bright stars that were not covered by \textit{Kepler}.

\subsection{Internal planetary structures}
\textcolor{blue}{ }

\noindent
We performed a Bayesian analysis in order to infer the possible interior structures of the three transiting planets\cite{Dorn15,Dorn17}. We assumed the planet to be made of four different layers: a central core made of iron and sulfur, a silicate mantle (containing Si, Mg, and Fe), a water layer, and a gas layer made of pure H and He. Compared to previous similar models\cite{Dorn15,Dorn17}, the physical models used here have been improved\cite{Mortier20}, namely with a new equation of state (EOS) for the water layer\cite{Haldemann_20}, and the EOS for the iron core\cite{Hakim} (which can also contain some sulfur). The EOS for the silicate mantle\cite{Sotin} depends on the mole fractions of Si, Mg, and Fe, and the gas envelope model\cite{LopezFortney14} gives the thickness of the gas envelope as a function of age, planetary mass, etc. In this analysis, we did not include the compression effect of the gas envelope onto the innermost layers of the planet (core, mantle, and water layer). The validity of this hypothesis can be checked \textit{a posteriori}, as the mass of gas is small for the three planets (see below).

\noindent
The transit and RV data provide measurements of the planetary radii and masses relative to those of the star. This introduced some correlation between the absolute planetary radii and masses, that we take into account by fitting with our model the \textit{planetary system} as a whole, and not each planet individually. For this, we assume that the planetary Si/Mg/Fe molar ratio for all planets is equal to the stellar one, and we fit directly the transit depths and radial velocity semi-amplitudes of the three planets at once. The data used by the model are therefore the stellar mass, radius, effective temperature, and age, as well as the chemical abundances of Fe, Mg, Si, and the planetary radial velocity semi-amplitudes, transit depths, and orbital periods. The prior distribution of the mass fractions of core, mantle, and water layer, which add up to one, is assumed to be uniform on the simplex (the surface defined by the sum of the three mass fractions equal to one). In addition, it is likely that the maximum water mass fraction of planets cannot be larger than the icy fraction inside planetary solid building blocks (whatever their nature - planetesimals or pebbles). We therefore assume that the mass fraction of water is less than 50\%\cite{Thiabaud,Marboeuf}. The prior of the gas mass is assumed to be uniform in log. The posterior distributions of the relevant internal structure parameters are shown in Supplementary Figures 12 to 14 for the three planets.  

\noindent
Under the assumed priors, the gas mass in the three planets shows strong variations between \hbox{planet b} which can be seen as a bare core, planet c with a gas mass (5\% and 95\% quantiles) of $0.13^{+0.103}_{-0.078}\:M_{\oplus}$ ($1.2_{-0.70}^{+0.91}$ weight percent wt\%), and \hbox{planet d} having a somewhat smaller gas mass of $0.058^{+0.069}_{-0.050}\:M_{\oplus}$ ($0.66_{-0.58}^{+0.79}$ wt\%). Since we derived the joint posterior probability of the three planetary internal structure parameters, we can easily estimate the probability that \hbox{planet d} has less gas than planet c, and we found this probability to be $\sim$90 \%. The water mass is found to be $0.57^{+0.60}_{-0.49}\:M_{\oplus}$ ($12.6_{-11.0}^{+14.7}$ wt\%) for the innermost planet b, $2.81^{+2.52}_{-2.52}\:M_{\oplus}$ ($25.01_{-22.4}^{+22.2}$ wt\%) for planet c, and $2.31^{+2.09}_{-2.13}\:M_{\oplus}$ ($26.8_{-23.7}^{+21.2}$ wt\%) for planet d. Finally, the radius of the high-Z part (iron core, mantle, and water layer) is similar for the two outer planets: $2.14^{+0.17}_{-0.20}\:R_{\oplus}$ for planet c and $2.04^{+0.19}_{-0.21}\:R_{\oplus}$ for \hbox{planet d}.

\subsection{Atmospheric evolution}
\textcolor{blue}{ }

\noindent
We constrained the initial atmospheric mass fraction of the three planets composing the $\nu^2$ Lupi system and the evolution of the rotation rate of the host star employing a slightly modified version of the tool presented by ref. \cite{kubyshkina2019a,kubyshkina2019b}. The algorithm models the evolution of planetary atmospheres accounting for atmospheric mass loss combining a model of the stellar high-energy (X-ray plus extreme ultraviolet; XUV) flux evolution, a model relating planetary parameters and atmospheric mass, and a model computing planetary atmospheric escape.

\noindent
The XUV flux of late-type stars (later than F5) out of the saturation regime depends on stellar mass and, more importantly, on rotation period\cite{wright2011}, which increases with time, but the evolution of the stellar rotation rate, and thus of the XUV emission, does not follow a unique path\cite{tu2015}. To account for the different rotation histories, the framework models the rotation period as a power law in age\cite{mamajek2008} normalised such that the computed rotation period at the present age is consistent with the current estimate\cite{kubyshkina2019a} of 23.8$\pm$3.1 days\cite{2019A&A...622A..37U}. The stellar XUV luminosity is then derived from the rotation period using scaling relations\cite{pizzolato2003,sanzForcada2011,wright2011}, while the evolution of the stellar bolometric luminosity, which drives the evolution of the equilibrium temperature of the planets, is obtained from interpolating a grid of stellar structure models\cite{choi2016,dotter2016}.

\noindent
To estimate the planetary atmospheric mass fraction as a function of mass, radius, and equilibrium temperature, we generated a grid of interior structure models as used in the previous section, and performed an average over the `hidden' parameters (stellar age, core radius, etc). Finally, the tool extracts the planetary mass-loss rates as a function of the system parameters by interpolating over a large grid of hydrodynamic upper atmosphere models\cite{kubyshkina2018a}. This approach has the advantage over the other commonly used analytical estimates, to appropriately and simultaneously account for both XUV-driven and core-powered mass loss\cite{kubyshkina2018a,kubyshkina2018b}. The key assumptions of the framework are that the planetary orbits do not change with time following the dispersal of the protoplanetary disk and that the planetary atmospheres are hydrogen-dominated.

\noindent
For each planet, the planetary atmospheric evolution calculations begin at an age of 5 Myr, which is the assumed age of the dispersal of the protoplanetary disk. At each time step, the framework then extracts the mass-loss rate from the grid employing the stellar flux and system parameters, and uses it to update the atmospheric mass fraction. This procedure is then repeated until the age of the system is reached or the planetary atmosphere has completely escaped. The free parameters of the algorithm are the initial atmospheric mass fraction at the time of the dispersal of the protoplanetary disk, and the index of the power law controlling the stellar rotation period (a proxy for the stellar XUV emission) within the first 2 Gyr. Afterwards, the relation between rotation period and age is modelled through a power law\cite{mamajek2008}, assuming the exponent 0.566.

\noindent
The free parameters are constrained by implementing the atmospheric evolution algorithm in a Bayesian framework employing a MCMC tool\cite{2017AJ....153....3C}. The framework uses the system parameters with their uncertainties as inputs (i.e. priors). It then computes millions of forward planetary evolutionary tracks, varying the input parameters according to the shape of the prior distributions, and varying the free parameters within pre-defined ranges, fitting the current planetary atmospheric mass fractions obtained as described in the previous section. The fit is done for the three planets simultaneously. The results are posterior distributions of the free parameters, which are the rotation period of the star when it was young and the planetary initial atmospheric mass fractions. The modifications to the original tool are fitting for the planetary atmospheric mass fractions, instead of the planetary radii, and employing the stellar rotation period as step parameter for the MCMC algorithm, instead of the index of the power law controlling the stellar rotation period within the first 2 Gyr. The former modification enables the code to be more accurate by avoiding to continuously convert the atmospheric mass fraction into planetary radius, given the other system parameters. The latter modification avoids biasing the MCMC calculations towards faster rotating stars.

\noindent
Supplementary Figure 15 shows the resulting posterior probability distribution functions (PDFs) for the rotation period of the host star at the age of 150 Myr and for the initial atmospheric mass fraction of each of the three planets. In this Figure, the stellar rotation period at an age of 150 Myr is also put in comparison to the distribution obtained from stars with masses between 0.75 and 1 $M_{\odot}$ members of open clusters of comparable age\cite{johnstone2015}. Our result suggests that $\nu^2$ Lupi evolved as a medium rotator, with a most probable rotation period at an age of 150\,Myr ranging between 1 and 10 days and peaking at about 7 days, in agreement with the rotation period observed for most of the stars member of open clusters of similar age.

\noindent
The posterior distribution obtained for the initial atmospheric mass fraction of $\nu^2$ Lupi\,b is flat, evidencing that the planet has completely lost its primary hydrogen-dominated atmosphere at some unknown point in time throughout the evolution. Therefore, the framework is unable to identify how much atmosphere there was when the planet stopped accreting. For both $\nu^2$ Lupi\,c and \hbox{$\nu^2$ Lupi\,d}, the posterior distribution of the initial atmospheric mass fraction presents one strong, rather narrow peak close to the one that is obtained for the current atmospheric mass fraction through the interior structure modelling (see previous section). These results suggest that both planets experienced little atmospheric evolution through mass loss. This indicates that both planets accreted only a small atmosphere and that their current low gas content is thus likely of primordial origin. On the basis of the evolution simulations, we estimate the current mass-loss rates of \hbox{$\nu^2$ Lupi\,c} and \hbox{$\nu^2$ Lupi\,d} to be of the order of 3.2$\times$10$^8$ and 1.8$\times$10$^7$\,g/s, respectively.

\subsection{Potential of the system for atmospheric characterization}
\textcolor{blue}{ }

\noindent
To assess quantitatively the potential of the system for atmospheric characterization, we used the Transmission Spectroscopy Metric (TSM) of ref. \cite{2018PASP..130k4401K}. This metric is proportional to the expected transmission spectroscopy signal-to-noise ratio (SNR) and defined as
\begin{align*}
    \mathrm{TSM} &= \mathrm{Scale\:factor} \times \frac{R_\mathrm{p}^3\: T_{\mathrm{eq}}}{M_{\mathrm{p}} \: R_{\star}^2} \times 10^{-m_{\mathrm{J}}/5}
\end{align*}
where $R_{\mathrm{p}}$ is the radius of the planet, $M_{\mathrm{p}}$ is its mass, $T_{\mathrm{eq}}$ is its equilibrium temperature, $R_{\star}$ is the radius of the star, and $m_{\mathrm{J}}$ is its apparent magnitude in the $J$-band. The scale factor depends on the radius of the planet and allows a one-to-one scaling between the TSM values and the SNRs estimated by ref. \cite{2018PASP..130d4401L} assuming 10 hours of observations with the Near InfraRed Imager and Slitless Spectrograph (NIRISS) aboard the \textit{James Webb Space Telescope (JWST)}. Using the system parameters derived from our global analysis (Table~\ref{table_glob}), we obtained TSM values of 125, 214, and 117, for $\nu^2$ Lupi b, c, and d, respectively. 
To provide context, Figure~\ref{fig:diagrams}d compares these TSM values with those of the currently known population of small ($R_{\mathrm{p}} < 4\:R_{\mathrm{\oplus}}$) transiting exoplanets as a function of their equilibrium temperature ($T_{\mathrm{eq}}$). The size of the symbols is proportional to the host star effective temperature. To identify the top atmospheric characterization targets among the exoplanet population, ref. \cite{2018PASP..130k4401K} recommends a threshold of 92 for planets with $1.5\: R_{\mathrm{\oplus}} < R_{\mathrm{p}} < 2.75\:R_{\mathrm{\oplus}}$, such as $\nu^2$ Lupi b and d, and a threshold of 84 for planets with $2.75\:R_{\mathrm{\oplus}} < R_{\mathrm{p}} < 4\:R_{\mathrm{\oplus}}$, such as \hbox{$\nu^2$ Lupi c}. All three $\nu^2$ Lupi planets are above these suggested thresholds, thus opening up promising perspectives for comparative atmospheric studies. 
We note that the TSM is only intended as a general metric for the ranking of transmission spectroscopy targets and that atmospheric observations of the $\nu^2$ Lupi planets may turn out to be challenging in practice. With a $K$-band magnitude of 4.16, $\nu^2$ Lupi is close to $JWST$ saturation limit for spectroscopy of $K\sim4$\cite{2014PASP..126.1134B}. Still, higher efficiency readout modes for observations of bright stars are currently being investigated\cite{2018ApJ...856L..34B}. In particular, a new mode\cite{2017PASP..129a5001S} was proposed for the Near InfraRed Camera (NIRCam) that would allow one to measure spectra of targets up to $K\sim1-2$ between 1 and 2 $\mu$m. 

\noindent
With its rocky and mostly dry composition, $\nu^2$ Lupi b might not be an ideal target for atmospheric characterization. Still, its high temperature ($\sim$900 K) opens the interesting possibility that its surface is molten and sustains a secondary atmosphere in equilibrium with the underlying magma\cite{2015ApJ...801..144I}. High-resolution ultraviolet spectroscopy with the \textit{Hubble Space Telescope} (\textit{HST}) could be used to detect the strong electronic transitions from metal effluents in this envelope..

\noindent
In contrast, the nature of $\nu^2$ Lupi c makes it a particularly promising target. Among the three planets, this is the one with the lowest bulk density. Measuring the hydrogen and helium content of its gas envelope, and its mass loss, would bring useful constraints to simulations of the planet evolution. High-resolution ground-based spectroscopy in the near-infrared will allow measuring the absorption lines from metastable helium in the upper atmosphere\cite{2018Sci...362.1384A,2018Sci...362.1388N}. Interestingly, \hbox{$\nu^2$ Lupi c} is in similar irradiation conditions as GJ 3470 b (Figure~\ref{fig:diagrams}), a warm Neptune that was found to be dramatically evaporating\cite{2018A&A...620A.147B} (at a rate of about $10^{10}$ g/s) and whose atmosphere has already been intensively studied using both space-\cite{2018A&A...620A.147B,2019NatAs...3..813B} and ground-based\cite{2020A&A...638A..61P} facilities. While the smaller present-day size of \hbox{$\nu^2$ Lupi c} probably makes it less efficient at capturing the stellar energy and evaporate, its lower density could favor the formation of a large exosphere of neutral hydrogen, sustained by the photodissociation of water from its massive reservoir\cite{2004ApJ...605L..65J}. At only 14.7 parsecs, the Lyman-$\alpha$ line of a G-type star like \hbox{$\nu^2$ Lupi} will show reduced absorption by the interstellar medium and could readily be used to search for the absorption signature of this exosphere with \textit{HST}\cite{2015Natur.522..459E}.

\noindent
At lower temperatures ($T_{\mathrm{eq}}<500$ K), the only target more favorable than $\nu^2$ Lupi d according to the TSM is L 98-59 d, a low-density super-Earth transiting an M-dwarf star\cite{2019A&A...629A.111C}. Despite a somewhat small gas envelope, the long-period and bright host star of $\nu^2$ Lupi d make it a unique target to probe a low-temperature atmosphere around a Sun-like star. Water absorption bands could be searched for in the near future with the \textit{JWST}, or with future ground-based Extremely Large Telescopes using the cross-correlation technique\cite{2010Natur.465.1049S,2017A&A...606A.144A}. Meanwhile, searching for the signature of helium in the near-infrared (using e.g. the upcoming NIRPS spectrograph in the Southern hemisphere) would allow disentangling between a water-dominated and a H/He envelope. We note, however, that the long and rare transits of $\nu^2$ Lupi d will be an important limitation for the study of this planet from the ground.

\subsection{Prospects for moons or rings around $\nu^2$ Lupi d}
\textcolor{blue}{ }

\noindent
The possibility for a planet to have rings depends primarily on the size of its Roche radius $R_r=2.456\:R_p (\rho_p/\rho_r)^{1/3}$, where $R_p$ is the planetary radius, and $\rho_p$ and $\rho_r$ are the planet and ring material density, respectively. Since planet d has an equilibrium temperature above 400 K, hypothetical rings should be water-free and presumably silicate-rich, with $\rho_r \simeq$ 3 g/cm$^3$. With the parameters in Table \ref{table_glob}, we get $R_r=2.4\:R_p$ which leaves open the possibility for planet d to have enough space for rings in its Roche zone. Either the rings could be faint and dusty (with optical depth $\ll 1$, like those of Jupiter), in which case their presence would not affect the measurement of the planet's radius, or they could be dense and opaque (with optical depth $>1$), which would then lead to an overestimate of the planet's radius. Rings can be detected in transit mostly during the ingress and egress phases, which were unfortunately not covered in the \textit{CHEOPS} dataset presented here. If the planet is tidally locked, the rings may be torqued toward the planet's orbital plane beyond about \hbox{1 $R_p$} making their detection in transit difficult as they would then be seen edge-on\cite{Tremaine_2009, Charnoz_2018}. However, our tidal evolution simulations (see above) suggest that \hbox{$\nu^2$ Lupi d} may not be tidally locked yet. In our Solar System, whereas the terrestrial planets may theoretically have rings (following the same arguments as above), they do not. The absence of rings around terrestrial planets is still not understood but could be attributed to the Poynting Robertson effect, by which micrometer-sized grains spiral down to the planet\cite{Burns_2001} by interacting with the star light. In that case, the relative proximity of planet d to its host star might prevent the long-term survival of dusty rings. Other putative removal processes are plasma-drag, or meteoritic or ion sputtering\cite{Charnoz_2018}. Future \textit{CHEOPS} observations with a better coverage of the ingress and egress phases will make it possible to further investigate these questions.

\noindent
Satellites orbiting a planet are subject to rapid tidal evolution, leading to their radial migration due to the exchange of angular momentum with the host planet. The region inside which satellites may be gravitationally bound to a planet is a fraction of the Hill Radius ($R_h=a (M_p/3M_*)^{1/3}$), with $M_p$ the mass of the planet, $a$ its orbital semi-major axis, and $M_{\star}$ the stellar mass. For \hbox{$\nu^2$ Lupi d}, we get $R_h=84 \:R_p$. Depending on whether the planet is tidally locked or not, its synchronous radius may be as far as \hbox{125 $R_p$}, implying that everywhere inside the planet's Hill sphere, any satellite should migrate toward the planet, at a increasingly faster rate as the satellite gets closer to the planet. A simple analysis of tidal evolution with averaged $da/dt$ equations\cite{Goldreich_1966} using the dissipation $Q$ and $k2$ Love number of the planet, shows that over 12 Gyr (estimated age of the system, see Table~\ref{table_glob}), a \hbox{3000 kg/m$^3$} satellite with a 1000 km radius may survive if it orbits at $>10\:R_p$ and the planet's $Q \ge 100$. A bigger satellite with a 3000 km radius may survive only if it orbits at $>30\:R_p$. On the other hand, satellites smaller than 100 km may survive even as close as a few $R_p$ and if $Q>1$. However, $Q$ is a poorly constrained quantity. It may be in the range of 1-10 for terrestrial planets and $>1000$ for gas giant planets. For a low-density super-Earth like $\nu^2$ Lupi d, $Q$ is unknown, but presumably between the terrestrial and gas-giant values. Based on the exquisite photometric precision of the current dataset, observing the transit of the whole planet's Hill sphere with \textit{CHEOPS} could allow the detection of satellites down to the Mars-size regime.

\end{methods}

\begin{addendum}

\item[Data availability] The \textit{CHEOPS} light curves used in this work will be made available for download at the CDS (Centre de donn\'ees astronomiques de Strasbourg). We will provide both the raw and detrended light curves. All other data that support the plots within this paper and other findings of this study are available from the corresponding author upon reasonable request.

\item[Code availability] The \textit{CHEOPS} DRP is built over several public python libraries, such as \texttt{astropy}\cite{astropy:2013,astropy:2018}, \texttt{numpy}\cite{numpy}, and \texttt{scipy}\cite{scipy}. The \textit{TESS} light curve was extracted using the \texttt{lightkurve}\cite{Lightkurve2018} open-source python package. The data analysis was performed using the \texttt{juliet} python library, which is also publicly available. The figures were produced using the \texttt{matplotlib}\cite{matplotlib} and \texttt{corner}\cite{corner} open-source python modules. The codes used in this work are available upon reasonable request from the corresponding author.

\end{addendum}

\supplement

\pagebreak

\begin{table*}
\centering
\begin{tabular}{cccccc}
\hline
\textbf{Visit} & \textbf{Start date and time (UTC)} & \textbf{Duration (h)} & \textbf{Data points} & \textbf{Efficiency (\%)} & \textbf{Planet(s)} \\
\hline
1 & 2020-04-04T15:13:16.9 & 11.56 & 526 & 55.9 & b \\
2 & 2020-04-14T16:21:58.1 & 10.96 & 538 & 60.3 & c \\
3 & 2020-04-16T03:59:45.7 & 11.60 & 545 & 57.7 & b \\
4 & 2020-04-27T18:01:43.2 & 12.83 & 611 & 58.5 & b \\
5 & 2020-06-08T21:33:50.9 & 11.65 & 504 & 53.1 & c,d \\
6 & 2020-07-06T10:40:06.7 & 11.56 & 466 & 49.5 & b,c \\
\hline
\end{tabular}
\caption{\textbf{Supplementary Table 1.} Log of \textit{CHEOPS} observations of $\nu^2$ Lupi.}   
\label{observing_log}
\end{table*}

\nopagebreak

\begin{table*}[!]
\centering
\begin{tabular}{cc}
\hline
\textbf{Elements} & \textbf{Abundances (dex)} \\
\hline
$[\mathrm{Fe/H}]$ & $-0.34 \pm 0.04$  \\
$[\mathrm{Mg/H}]$ & $-0.14 \pm 0.05$  \\
$[\mathrm{Al/H}]$ & $-0.13 \pm 0.02$  \\
$[\mathrm{Si/H}]$ & $-0.23 \pm 0.03$ \\
$[\mathrm{Ca/H}]$ & $-0.24 \pm 0.05$ \\
$[\mathrm{Ti/H}]$ & $-0.17 \pm 0.04$ \\
$[\mathrm{Cr/H}]$ & $-0.32 \pm 0.05$ \\
$[\mathrm{Ni/H}]$ & $-0.35 \pm 0.02$ \\
$[\mathrm{\alpha/H}]$ & $-0.19 \pm 0.02$ \\
\hline
\end{tabular}
\caption{\textbf{Supplementary Table 2.} Detailed stellar abundances of $\nu^2$ Lupi. \label{abundances}} 
\end{table*}

\clearpage

\begin{scriptsize}
\begin{longtable}{lcr}
\hline
\textbf{Parameters} & \textbf{Priors} & \textbf{Description} \\
\hline
\bf{Physical parameters} & & \\
$P_b$ (d) & $\mathcal{U}$(11.53, 11.63) & Orbital period of planet b\\
$T_{0,b}$ ($\mathrm{BJD_{TDB}}$) & $\mathcal{U}$(2458944.28, 2458944.44) & Mid-transit time of planet b\\
$r_{1,b}$ & $\mathcal{U}$(0, 1) & Parametrization for $p$ and $b$ for planet b\\
$r_{2,b}$ & $\mathcal{U}$(0, 1) & Parametrization for $p$ and $b$ for planet b\\
$K_{b}$ (m/s) & $\mathcal{U}$(0, 10) & RV semi-amplitude for planet b\\
$\sqrt{e_b}\:\mathrm{sin}\,\omega_b$ & $\mathcal{U}$(-1, 1) or fixed to 0 for the circular model & Parametrization for $e$ and $\omega$ for planet b\\
$\sqrt{e_b}\:\mathrm{cos}\,\omega_b$ & $\mathcal{U}$(-1, 1) or fixed to 0 for the circular model & Parametrization for $e$ and $\omega$ for planet b\\
$P_c$ (d) & $\mathcal{U}$(27.54, 27.64) & Orbital period of planet c\\
$T_{0,c}$ ($\mathrm{BJD_{TDB}}$) & $\mathcal{U}$(2458954.35, 2458954.47) & Mid-transit time of planet c\\
$r_{1,c}$ & $\mathcal{U}$(0, 1) & Parametrization for $p$ and $b$ for planet c\\
$r_{2,c}$ & $\mathcal{U}$(0, 1) & Parametrization for $p$ and $b$ for planet c\\
$K_{c}$ (m/s) & $\mathcal{U}$(0, 10) & RV semi-amplitude for planet c\\
$\sqrt{e_c}\:\mathrm{sin}\,\omega_c$ & $\mathcal{U}$(-1, 1) or fixed to 0 for the circular model & Parametrization for $e$ and $\omega$ for planet c\\
$\sqrt{e_c}\:\mathrm{cos}\,\omega_c$ & $\mathcal{U}$(-1, 1) or fixed to 0 for the circular model & Parametrization for $e$ and $\omega$ for planet c\\
$P_d$ (d) & $\mathcal{U}$(105, 110) & Orbital period of planet d\\
$T_{0,d}$ ($\mathrm{BJD_{TDB}}$) & $\mathcal{U}$(2459009.6, 2459010.0) & Mid-transit time of planet d\\
$r_{1,d}$ & $\mathcal{U}$(0, 1) & Parametrization for $p$ and $b$ for planet d\\
$r_{2,d}$ & $\mathcal{U}$(0, 1) & Parametrization for $p$ and $b$ for planet d\\
$K_{d}$ (m/s) & $\mathcal{U}$(0, 10) & RV semi-amplitude for planet d\\
$\sqrt{e_d}\:\mathrm{sin}\,\omega_d$ & $\mathcal{U}$(-1, 1) or fixed to 0 for the circular model & Parametrization for $e$ and $\omega$ for planet d\\
$\sqrt{e_d}\:\mathrm{cos}\,\omega_d$ & $\mathcal{U}$(-1, 1) or fixed to 0 for the circular model & Parametrization for $e$ and $\omega$ for planet d\\
$\rho_\star$ (kg/m$^{3}$) & $\mathcal{N}$(1035, 76$^2$) & Stellar density\\
$q_{1,TESS}$ & $\mathcal{U}$(0, 1) & Quadratic limb-darkening parametrization for \textit{TESS}\\
$q_{2,TESS}$ & $\mathcal{U}$(0, 1) & Quadratic limb-darkening parametrization for \textit{TESS}\\
$q_{1,CHEOPS}$ & $\mathcal{U}$(0, 1) & Quadratic limb-darkening parametrization for \textit{CHEOPS}\\
$q_{2,CHEOPS}$ & $\mathcal{U}$(0, 1) & Quadratic limb-darkening parametrization for \textit{CHEOPS}\\
\hline
\bf{Nuisance parameters} & & \\
$\sigma_{TESS}$ (ppm) & $\mathcal{J}$(1, 1000); $\mathcal{N}$(122.7, 1.1$^2$) & Extra jitter term for \textit{TESS}\\
$M_{TESS}$ & $\mathcal{N}$(0, 0.1$^2$); $\mathcal{N}$(-0.000004, 0.000003$^2$) & Relative flux offset for \textit{TESS}\\
$\sigma_{GP,TESS}$ (ppm) & $\mathcal{J}$(0.1, 10$^4$); $\mathcal{N}$(31, 2$^2$) & Amplitude of the Matérn GP for \textit{TESS}\\
$\rho_{GP,TESS}$ & $\mathcal{J}$(10$^{-4}$, 10$^4$); $\mathcal{N}$(0.055, 0.010$^2$) & Length-scale of the Matérn GP for \textit{TESS}\\
$\sigma_{CHEOPS1}$ (ppm) & $\mathcal{J}$(1, 1000); $\mathcal{N}$(26.3, 1.9$^2$) & Extra jitter term for the 1$^{\rm{st}}$ \textit{CHEOPS} visit\\
$M_{CHEOPS1}$ & $\mathcal{N}$(0, 0.1$^2$); $\mathcal{N}$(-0.000027, 0.000061$^2$) & Relative flux offset for the 1$^{\rm{st}}$ \textit{CHEOPS} visit\\
$\theta_{0,CHEOPS1}$ & $\mathcal{U}$(-1, 1); $\mathcal{N}$(-0.000040, 0.000088$^2$) & sin($\phi$) regression coefficient for the 1$^{\rm{st}}$ \textit{CHEOPS} visit\\
$\theta_{1,CHEOPS1}$ & $\mathcal{U}$(-1, 1); $\mathcal{N}$(-0.000092, 0.000062$^2$) & cos($\phi$) regression coefficient for the 1$^{\rm{st}}$ \textit{CHEOPS} visit\\
$\theta_{2,CHEOPS1}$ & $\mathcal{U}$(-1, 1); $\mathcal{N}$(0.000059, 0.000064$^2$) & sin(2$\phi$) regression coefficient for the 1$^{\rm{st}}$ \textit{CHEOPS} visit\\
$\theta_{3,CHEOPS1}$ & $\mathcal{U}$(-1, 1); $\mathcal{N}$(-0.000081, 0.000031$^2$) & cos(2$\phi$) regression coefficient for the 1$^{\rm{st}}$ \textit{CHEOPS} visit\\
$\theta_{4,CHEOPS1}$ & $\mathcal{U}$(-1, 1); $\mathcal{N}$(0.000048, 0.000018$^2$) & sin(3$\phi$) regression coefficient for the 1$^{\rm{st}}$ \textit{CHEOPS} visit\\
$\theta_{5,CHEOPS1}$ & $\mathcal{U}$(-1, 1); $\mathcal{N}$(-0.000019, 0.000030$^2$) & cos(3$\phi$) regression coefficient for the 1$^{\rm{st}}$ \textit{CHEOPS} visit\\
$\sigma_{CHEOPS2}$ (ppm) & $\mathcal{J}$(1, 1000); $\mathcal{N}$(26.5, 2.0$^2$) & Extra jitter term for the 2$^{\rm{nd}}$ \textit{CHEOPS} visit\\
$M_{CHEOPS2}$ & $\mathcal{N}$(0, 0.1$^2$); $\mathcal{N}$(0.000115, 0.000045$^2$) & Relative flux offset for the 2$^{\rm{nd}}$ \textit{CHEOPS} visit\\
$\theta_{0,CHEOPS2}$ & $\mathcal{U}$(-1, 1); $\mathcal{N}$(0.000149, 0.000055$^2$) & sin($\phi$) regression coefficient for the 2$^{\rm{nd}}$ \textit{CHEOPS} visit\\
$\theta_{1,CHEOPS2}$ & $\mathcal{U}$(-1, 1); $\mathcal{N}$(-0.000199, 0.000059$^2$) & cos($\phi$) regression coefficient for the 2$^{\rm{nd}}$ \textit{CHEOPS} visit\\
$\theta_{2,CHEOPS2}$ & $\mathcal{U}$(-1, 1); $\mathcal{N}$(0.000189, 0.000053$^2$) & sin(2$\phi$) regression coefficient for the 2$^{\rm{nd}}$ \textit{CHEOPS} visit\\
$\theta_{3,CHEOPS2}$ & $\mathcal{U}$(-1, 1); $\mathcal{N}$(-0.000013, 0.000017$^2$) & cos(2$\phi$) regression coefficient for the 2$^{\rm{nd}}$ \textit{CHEOPS} visit\\
$\theta_{4,CHEOPS2}$ & $\mathcal{U}$(-1, 1); $\mathcal{N}$(0.000044, 0.000021$^2$) & sin(3$\phi$) regression coefficient for the 2$^{\rm{nd}}$ \textit{CHEOPS} visit\\
$\theta_{5,CHEOPS2}$ & $\mathcal{U}$(-1, 1); $\mathcal{N}$(0.000043, 0.000021$^2$) & cos(3$\phi$) regression coefficient for the 2$^{\rm{nd}}$ \textit{CHEOPS} visit\\
$\sigma_{CHEOPS3}$ (ppm) & $\mathcal{J}$(1, 1000); $\mathcal{N}$(26.5, 1.9$^2$) & Extra jitter term for the 3$^{\rm{rd}}$ \textit{CHEOPS} visit\\
$M_{CHEOPS3}$ & $\mathcal{N}$(0, 0.1$^2$); $\mathcal{N}$(-0.000029, 0.000008$^2$) & Relative flux offset for the 3$^{\rm{rd}}$ \textit{CHEOPS} visit\\
$\theta_{0,CHEOPS3}$ & $\mathcal{U}$(-1, 1); $\mathcal{N}$(-0.000039, 0.000011$^2$) & sin($\phi$) regression coefficient for the 3$^{\rm{rd}}$ \textit{CHEOPS} visit\\
$\theta_{1,CHEOPS3}$ & $\mathcal{U}$(-1, 1); $\mathcal{N}$(-0.000003, 0.000011$^2$) & cos($\phi$) regression coefficient for the 3$^{\rm{rd}}$ \textit{CHEOPS} visit\\
$\sigma_{CHEOPS4}$ (ppm) & $\mathcal{J}$(1, 1000); $\mathcal{N}$(38.1, 3.4$^2$) & Extra jitter term for the 4$^{\rm{th}}$ \textit{CHEOPS} visit\\
$M_{CHEOPS4}$ & $\mathcal{N}$(0, 0.1$^2$); $\mathcal{N}$(-0.000099, 0.000014$^2$) & Relative flux offset for the 4$^{\rm{th}}$ \textit{CHEOPS} visit\\
$\theta_{0,CHEOPS4}$ & $\mathcal{U}$(-1, 1); $\mathcal{N}$(-0.000048, 0.000012$^2$) & sin($\phi$) regression coefficient for the 4$^{\rm{th}}$ \textit{CHEOPS} visit\\
$\theta_{1,CHEOPS4}$ & $\mathcal{U}$(-1, 1); $\mathcal{N}$(0.000084, 0.000022$^2$) & cos($\phi$) regression coefficient for the 4$^{\rm{th}}$ \textit{CHEOPS} visit\\
$\theta_{2,CHEOPS4}$ & $\mathcal{U}$(-1, 1); $\mathcal{N}$(0.000003, 0.000014$^2$) & sin(2$\phi$) regression coefficient for the 4$^{\rm{th}}$ \textit{CHEOPS} visit\\
$\theta_{3,CHEOPS4}$ & $\mathcal{U}$(-1, 1); $\mathcal{N}$(0.000024, 0.000013$^2$) & cos(2$\phi$) regression coefficient for the 4$^{\rm{th}}$ \textit{CHEOPS} visit\\
$\sigma_{CHEOPS5}$ (ppm) & $\mathcal{J}$(1, 1000); $\mathcal{N}$(33.3, 3.9$^2$) & Extra jitter term for the 5$^{\rm{th}}$ \textit{CHEOPS} visit\\
$M_{CHEOPS5}$ & $\mathcal{N}$(0, 0.1$^2$); $\mathcal{N}$(-0.000004, 0.000013$^2$) & Relative flux offset for the 5$^{\rm{th}}$ \textit{CHEOPS} visit\\
$\theta_{0,CHEOPS5}$ & $\mathcal{U}$(-1, 1); $\mathcal{N}$(0.000003, 0.000011$^2$) & sin($\phi$) regression coefficient for the 5$^{\rm{th}}$ \textit{CHEOPS} visit\\
$\theta_{1,CHEOPS5}$ & $\mathcal{U}$(-1, 1); $\mathcal{N}$(0.000037, 0.000014$^2$) & cos($\phi$) regression coefficient for the 5$^{\rm{th}}$ \textit{CHEOPS} visit\\
$\sigma_{CHEOPS6}$ (ppm) & $\mathcal{J}$(1, 1000); $\mathcal{N}$(34.8, 3.9$^2$) & Extra jitter term for the 6$^{\rm{th}}$ \textit{CHEOPS} visit\\
$M_{CHEOPS6}$ & $\mathcal{N}$(0, 0.1$^2$); $\mathcal{N}$(-0.000062, 0.000011$^2$) & Relative flux offset for the 6$^{\rm{th}}$ \textit{CHEOPS} visit\\
$\theta_{0,CHEOPS6}$ & $\mathcal{U}$(-1, 1); $\mathcal{N}$(0.000021, 0.000017$^2$) & sin($\phi$) regression coefficient for the 6$^{\rm{th}}$ \textit{CHEOPS} visit\\
$\theta_{1,CHEOPS6}$ & $\mathcal{U}$(-1, 1); $\mathcal{N}$(0.000064, 0.000011$^2$) & cos($\phi$) regression coefficient for the 6$^{\rm{th}}$ \textit{CHEOPS} visit\\
$S0_{GP,CHEOPS}$ & $\mathcal{J}$($10^{-20}$, $10^{20}$);  $\mathcal{N}$(5.2 $10^{-12}$,(6.0 10$^{-13}$)$^2$) & Characteristic power of the SHO GP for \textit{CHEOPS}\\
$\omega 0_{GP,CHEOPS}$ & $\mathcal{J}$($10^{-5}$, $10^{5}$); $\mathcal{N}$(838, 60$^2$) & Characteristic frequency of the SHO GP for \textit{CHEOPS}\\
$\sigma_{\rm{HARPS-pre}}$ (m/s) & $\mathcal{J}$(0.01, 10); $\mathcal{N}$(1.34, 0.07$^2$) & Extra jitter term for HARPS pre-upgrade\\
$\mu_{\rm{HARPS-pre}}$ (m/s) & $\mathcal{U}$(-68720, -68700); $\mathcal{N}$(-68709.02, 0.09$^2$) & Systemic velocity for HARPS pre-upgrade\\
$\sigma_{\rm{HARPS-post}}$ (m/s) & $\mathcal{J}$(0.01, 10); $\mathcal{N}$(0.20, 0.58$^2$) & Extra jitter term for HARPS post-upgrade\\
$\mu_{\rm{HARPS-post}}$ (m/s) & $\mathcal{U}$(-68705, -68685); $\mathcal{N}$(-68695.10, 0.36$^2$) & Systemic velocity for HARPS post-upgrade\\
\hline
\caption{\textbf{Supplementary Table 3.} Priors used in our data analyses with \texttt{juliet}. $\mathcal{N}(\mu, \sigma^2)$ represents a normal distribution of mean $\mu$ and variance $\sigma^2$. $\mathcal{U}(a, b)$ represents a uniform distribution between $a$ and $b$. $\mathcal{J}(a, b)$ represents a Jeffrey's prior (i.e. a log-uniform distribution) between $a$ and $b$. For the nuisance parameters, two prior distributions are given. The first ones are the wide priors that we assumed for the preliminary individual analyses of each dataset. The second ones are the normal priors that we assumed in our global analysis (to aid convergence), based on the results from the preliminary individual analyses.}
\label{table_priors}
\end{longtable}
\end{scriptsize}

\clearpage

\begin{table*}[h]
\centering
\begin{tabular}{lc}
\hline
\hline
\textbf{Models} & \textbf{$\Delta$ ln $Z$} \\
\hline
\hline
\textit{Reference model:} & \\
3 planets, circular orbits for all planets, \textit{CHEOPS} single transit caused by d & $-$ \\
\hline
\textit{Tests for eccentricities:} & \\
Eccentric orbit for b, circular orbits for c and d & $-0.9$ \\
Eccentric orbit for c, circular orbits for b and d & $-1.7$ \\
Eccentric orbit for d, circular orbits for b and c & $-0.9$ \\
Eccentric orbits for all planets & $-3.6$ \\
\hline
\textit{Tests for a fourth planet and origin of the CHEOPS single transit:} & \\
4 planets ($P_{e}$=485d), \textit{CHEOPS} single transit caused by e, d not transiting & $-2.5$ \\
4 planets ($P_{e}$=485d), \textit{CHEOPS} single transit caused by d, e not transiting & $+0.6$ \\
\hline
\end{tabular}
\caption{\textbf{Supplementary Table 4.} Comparison of different models based on their Bayesian log evidence.  Based on these tests, we adopted the reference model as our nominal solution (3 planets with circular orbits, single \textit{CHEOPS} transit caused by planet d). \label{tab:evidences}} 
\end{table*}

\begin{table*}[!]
\begin{scriptsize}
\begin{center}
\begin{tabular}{ c | c | c | c | c }
\hline 
 \textbf{Parameters} &  \textbf{Ref. \cite{2020AJ....160..129K}} &  \textbf{\textit{TESS}} &  \textbf{\textit{CHEOPS}} &  \textbf{Global analysis} \\
 & (\textit{TESS}+RVs) & (+RVs) & (+RVs) & (\textit{TESS}+\textit{CHEOPS}+RVs) \\
\hline
\textbf{Planet b} & & & & \\[0.1cm]
$R_p/R_{\star}$ & $0.01343_{-0.00045}^{+0.00044}$ & $0.01323_{-0.00074}^{+0.00075}$ & $0.01439_{-0.00041}^{+0.00040}$ & $0.01428_{-0.00038}^{+0.00036}$  \\[0.1cm]
$R_{\mathrm{p}}$ ($R_{\oplus}$) & $1.482_{-0.056}^{+0.058}$ & $1.527 \pm 0.090$ & 1.661 $\pm$ 0.055 & $1.648_{-0.051}^{+0.052}$ \\[0.1cm]
$b$ ($R_{\star}$) & $0.41_{-0.20}^{+0.12}$ & $0.38_{-0.22}^{+0.15}$ & $0.48_{-0.16}^{+0.09}$ & $0.47_{-0.16}^{+0.09}$ \\[0.1cm]
$W$ (hours) & $3.936_{-0.108}^{+0.115}$ & $4.25_{-0.34}^{+0.25}$ & $3.95_{-0.08}^{+0.14}$ & $3.940_{-0.064}^{+0.103}$  \\[0.1cm]
$T_0\:(\mathrm{BJD_{TDB}} - 2,450,000)$ & $8631.7672_{-0.0022}^{+0.0023}$ & $8631.7723_{-0.0050}^{+0.0036}$ & $8944.3718_{-0.0026}^{+0.0016}$ & $8944.3724_{-0.0019}^{+0.0015}$  \\[0.1cm]
$P$ (days) & $11.57779_{-0.0011}^{+0.00091}$ & $11.57748_{-0.00124}^{+0.00103}$ & $11.57822_{-0.00061}^{+0.00045}$ & $11.57795_{-0.00014}^{+0.00009}$  \\[0.1cm]
$e$ & $0.079_{-0.053}^{+0.068}$ & $0.098_{-0.064}^{+0.076}$ & $0.066_{-0.045}^{+0.058}$ & $0.076_{-0.046}^{+0.047}$  \\[0.1cm]
\hline
\textbf{Planet c} & & & & \\[0.1cm]
$R_p/R_{\star}$ & 0.02363 $\pm$ 0.00052 & $0.0249_{-0.0010}^{+0.0012}$  & $0.02551_{-0.00051}^{+0.00055}$ & $0.02527_{-0.00050}^{+0.00046}$  \\[0.1cm]
$R_{\mathrm{p}}$ ($R_{\oplus}$) & $2.608_{-0.077}^{+0.078}$ & $2.87_{-0.13}^{+0.14}$ & $2.944_{-0.079}^{+0.084}$ & $2.918_{-0.077}^{+0.074}$ \\[0.1cm]
$b$ ($R_{\star}$) & $0.854_{-0.016}^{+0.013}$ & $0.874_{-0.013}^{+0.012}$ & $0.876 \pm 0.010$ & $0.873 \pm 0.010$ \\[0.1cm]
$W$ (hours) & $3.209_{-0.053}^{+0.058}$ & $3.24 \pm 0.11$ & $3.242_{-0.038}^{+0.041}$ & $3.254_{-0.033}^{+0.039}$ \\[0.1cm]
$T_0\:(\mathrm{BJD_{TDB}} - 2,450,000)$ & $8650.8947_{-0.0010}^{+0.0011}$ & $8650.8959_{-0.0012}^{+0.0013}$ & $8954.40959_{-0.00070}^{+0.00067}$  & $8954.40987_{-0.00054}^{+0.00051}$  \\[0.1cm]
$P$ (days) & $27.5909_{-0.0031}^{+0.0028}$ & $27.5911_{-0.0034}^{+0.0029}$ & $27.59255_{-0.00048}^{+0.00045}$ & 27.59220 $\pm$ 0.00011  \\[0.1cm]
$e$ & $0.037_{-0.026}^{+0.039}$ & $0.038_{-0.027}^{+0.042}$ & $0.022_{-0.016}^{+0.027}$ & $0.022_{-0.015}^{+0.026}$ \\[0.1cm]
\hline
\end{tabular}
\caption{\textbf{Supplementary Table 5.} Comparison between the results of different data analyses for some key transit parameters of the planets $\nu^2$ Lupi b and c. The second column presents the values reported by ref. \cite{2020AJ....160..129K} based on their analysis of the \textit{TESS} and RV data. In the third column, we report the results of our own analysis of this same dataset. The fourth column gives the results of our analysis of the six \textit{CHEOPS} visits together with the RVs, while the fifth column lists the results from our global analysis of all the data. We assumed eccentric orbits for these analyses, to allow a direct comparison with the results of ref. \cite{2020AJ....160..129K}. However, we note that circular orbits are actually favored for all the planets (see section ``Global data analysis'' and Table 1 for our nominal solution).
\label{tab:compa}} 
\end{center}
\end{scriptsize}
\end{table*}

\begin{figure*}
\vspace{-1cm}
\centerline{\includegraphics[width=17cm,height=23cm]{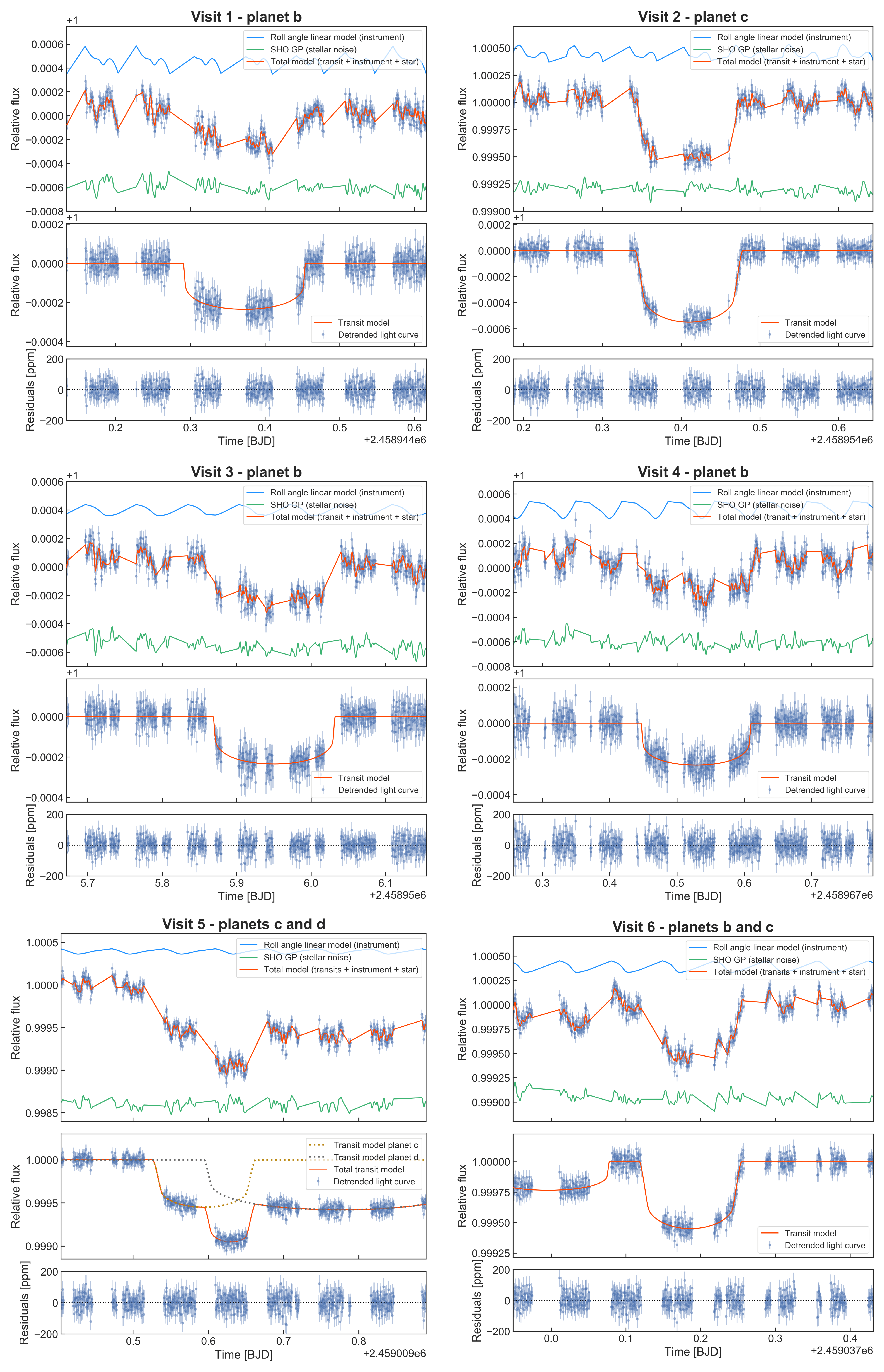}}
\vspace{-0.5cm}
\caption{\begin{small}\textbf{Supplementary Figure 1.} Individual \textit{CHEOPS} light curves. For each visit, the top panel shows the raw light curve (blue points with error bars), together with the best-fit instrumental noise (blue curve shifted vertically for visualisation), stellar noise (green curve shifted vertically), and total (transit + instrument + stellar noise, orange line) models. The second panel shows the light curve corrected from the best-fit instrumental and stellar noise models, together with the best-fit transit model. The bottom panel shows the residuals. The error bars correspond to the quadratic sum of the formal photometric errors and the fitted extra jitter term.\end{small} \label{fig:CHEOPS_LCs_indiv}}
\end{figure*}

\begin{figure*}
\centerline{\includegraphics[width=0.78\hsize]{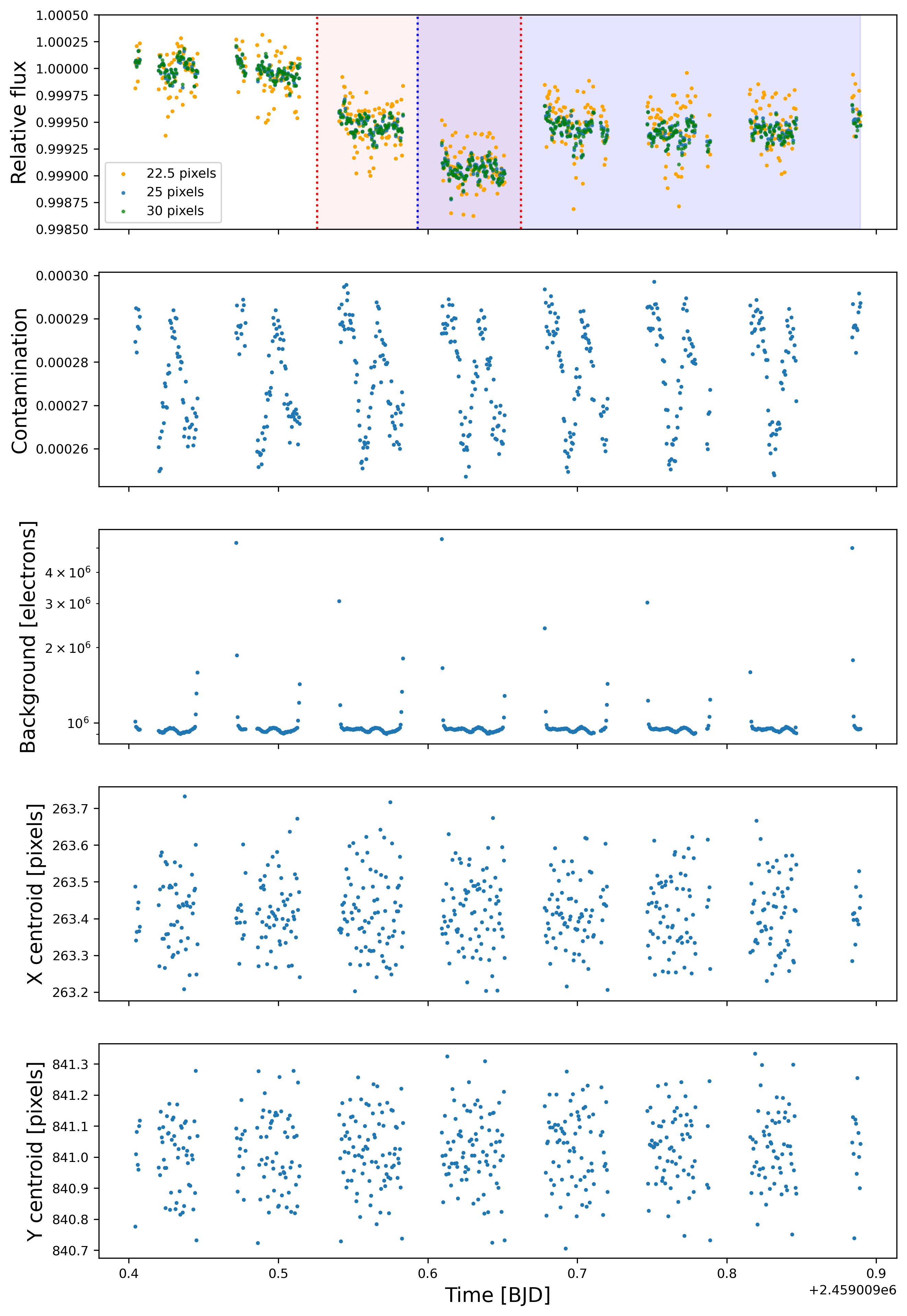}}
\vspace{-0.5cm}
\caption{\begin{small}\textbf{Supplementary Figure 2.} Fifth \textit{CHEOPS} visit. \textit{Upper panel:} Raw light curves obtained for three different photometric apertures. All the light curves show an unexpected $\sim$500 ppm transit-like flux drop which started during the targeted transit of planet c (transit window indicated as a red shaded region) and lasted for the rest of the visit (blue shaded region). This signal does not show any correlation with instrumental or environmental parameters, of which the main ones are shown in the other panels. \textit{Second panel:} Contamination from nearby stars entering the photometric aperture (25 pixels), relative to the target's flux. \textit{Third panel:} Background (in electrons). Note that it is higher right before/after Earth occultations due to stray light. \textit{Fourth and fifth panels:} $x$- and $y$- position of the target's PSF centroid on the CCD. 
\end{small} \label{fig:visit5_ext_param}}
\end{figure*}

\begin{figure*} 
\centerline{\includegraphics[width=0.5\hsize]{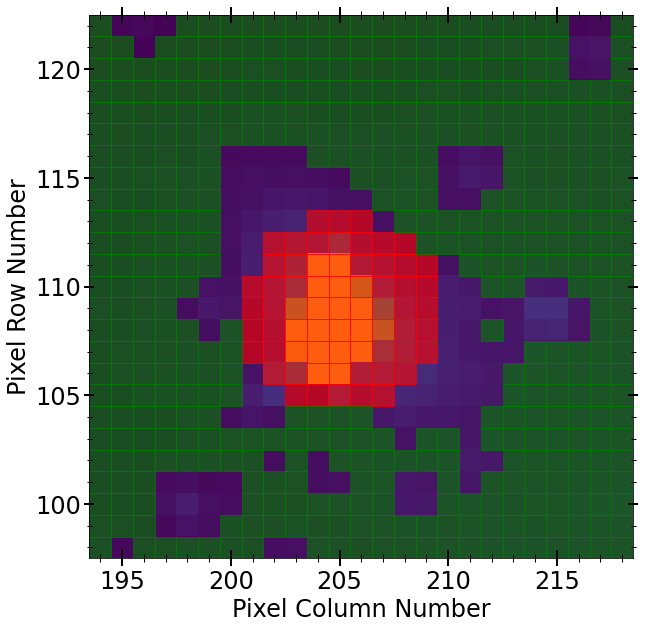}}
\caption{\textbf{Supplementary Figure 3.} Sample \textit{TESS} target pixel file and masks used to measure the target flux (red) and background (green). For the background, we used pixels with median fluxes lower than one standard deviation above the overall median of the target pixel file.}
\label{fig:TESS_aper}
\end{figure*}

\begin{figure*}
\vspace{-2cm}
\centerline{\includegraphics[width=0.87\hsize]{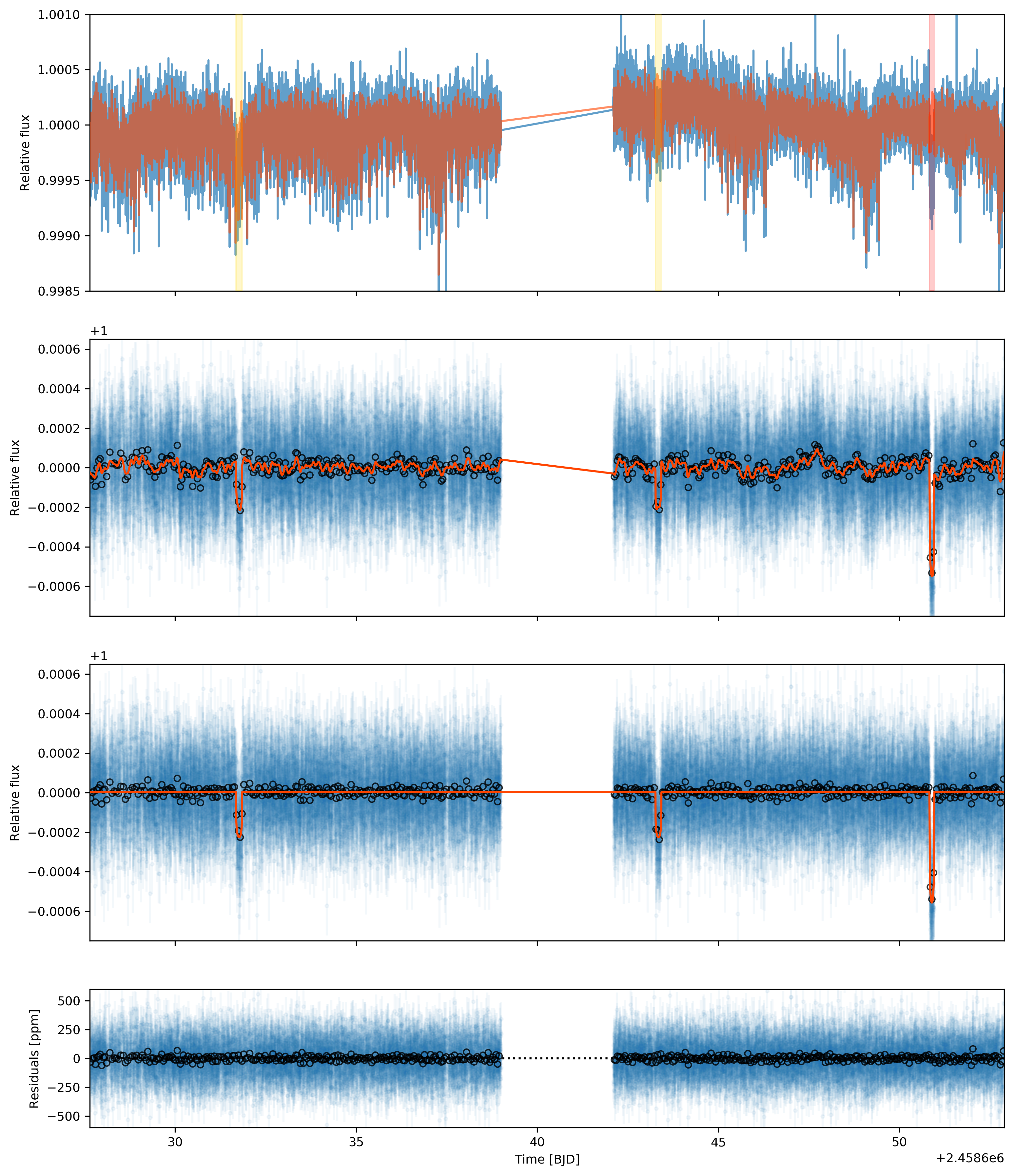}}
\caption{\begin{small}\textbf{Supplementary Figure 4.} \textit{TESS} photometry of $\nu^2$ Lupi. \textit{Upper panel:} Raw photometry (in blue) along with our best-fit linear regression model (in orange). The locations of the two transits of planet b and single transit of planet c are indicated as yellow and red shaded regions, respectively. \textit{Second panel:} Light curve (blue dots with error bars) corrected for most instrumental systematics that we used in our global analysis. The orange line shows our best-fit model which includes both the transits of planets b and c, and the GP model used to account for the remaining photometric variability. The black open circles show the light curve binned into 1-hour intervals. \textit{Third panel:} Light curve obtained after subtracting the GP component of our model, together with our best-fit transit model. \textit{Lower panel:} Corresponding residuals. For all panels, the error bars of the data points include the fitted extra jitter term added in quadrature. \end{small}
\label{fig:TESS}}
\end{figure*}

\begin{figure*} 
\centerline{\includegraphics[width=0.9\hsize]{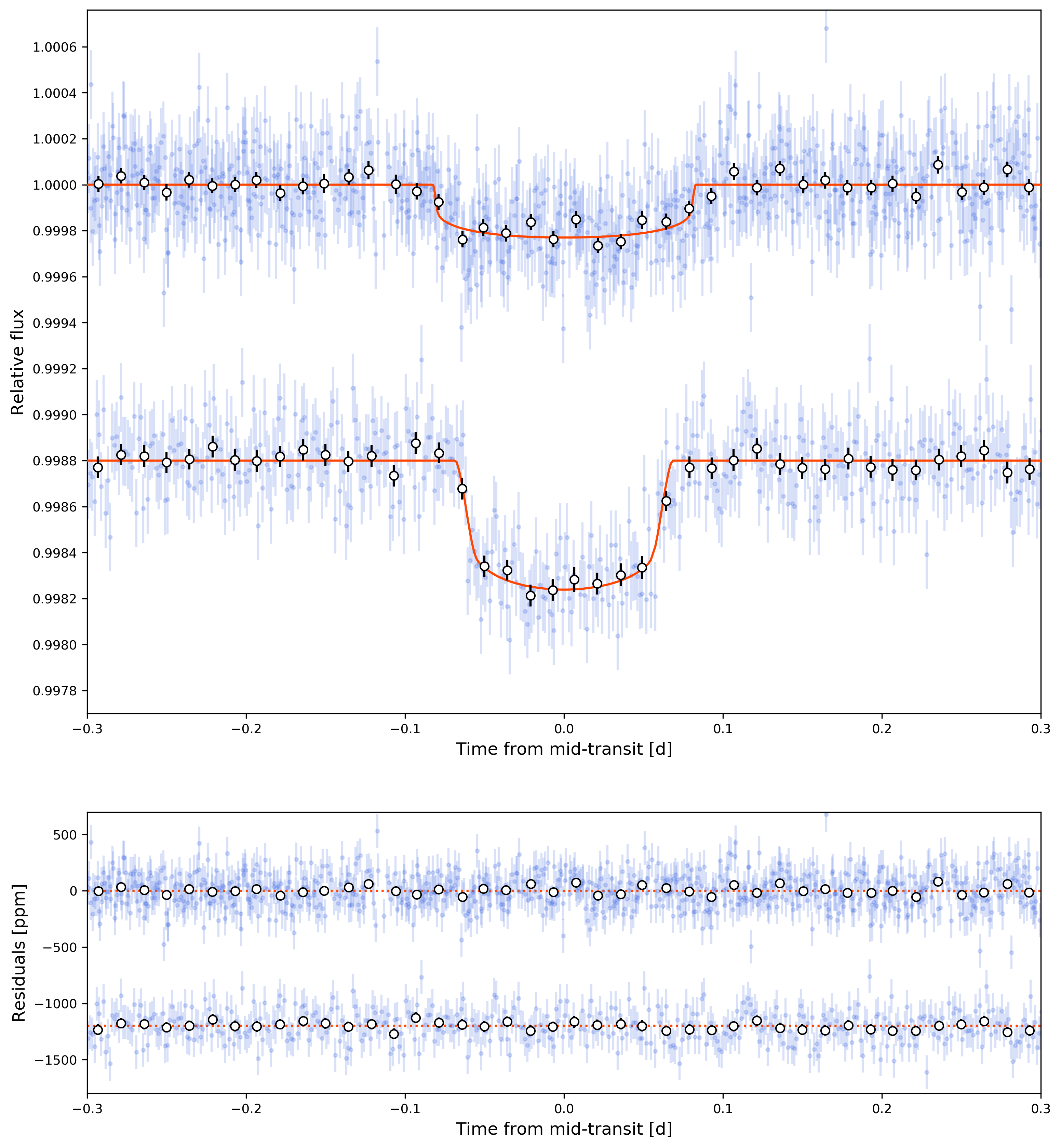}}
\caption{\textbf{Supplementary Figure 5.} \textit{Upper panel:} Corrected and phase-folded \textit{TESS} transit photometry of $\nu^2$ Lupi b (top) and c (bottom). The blue dots show the unbinned measurements, with error bars corresponding to the quadratic sum of the formal photometric errors and the fitted extra jitter term. Open circles show the light curves binned into 20-minute intervals, with error bars corresponding to the standard deviation of the data for each bin. \textit{Lower panel:} Corresponding residuals.\label{fig:TESS_folded}}
\end{figure*}

\begin{figure*} 
\vspace{-1cm}
\centerline{\includegraphics[width=0.75\hsize]{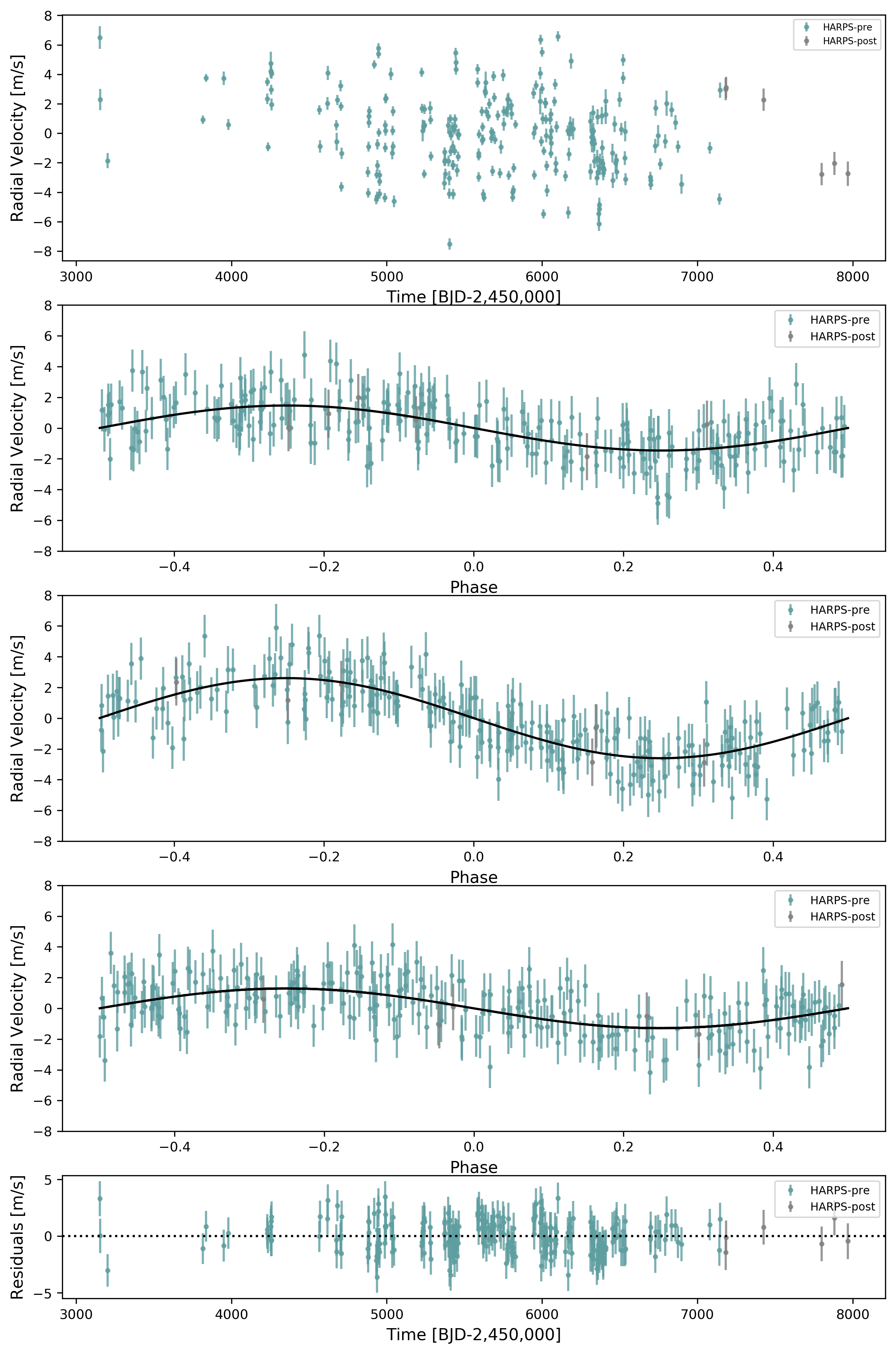}}
\caption{\begin{small}\textbf{Supplementary Figure 6.} HARPS radial velocities of $\nu^2$ Lupi used in our analysis. Measurements obtained before the instrument upgrade are shown in blue, while those obtained after are shown in grey. \textit{Upper panel:} The RVs are shown as a function of Barycentric Julian Date. The \textit{second}, \textit{third}, and \textit{fourth panels} show the measurements folded on the orbital periods of planets b, c, and d, respectively, together with the respective best-fit RV models (solid lines). For each of these panels, the RVs were corrected for the signals of the other two planets, so that only the signal of the considered planet is visible. \textit{Lower panel:} Residuals around the best-fit solution. For the upper panel, the error bars are the formal measurement uncertainties. For the other panels, the error bars are the quadratic sum of the measurement uncertainties and the fitted extra jitter term. \end{small}
\label{fig:RVs}}
\end{figure*}

\begin{figure*} 
\centerline{\includegraphics[width=1.0\hsize]{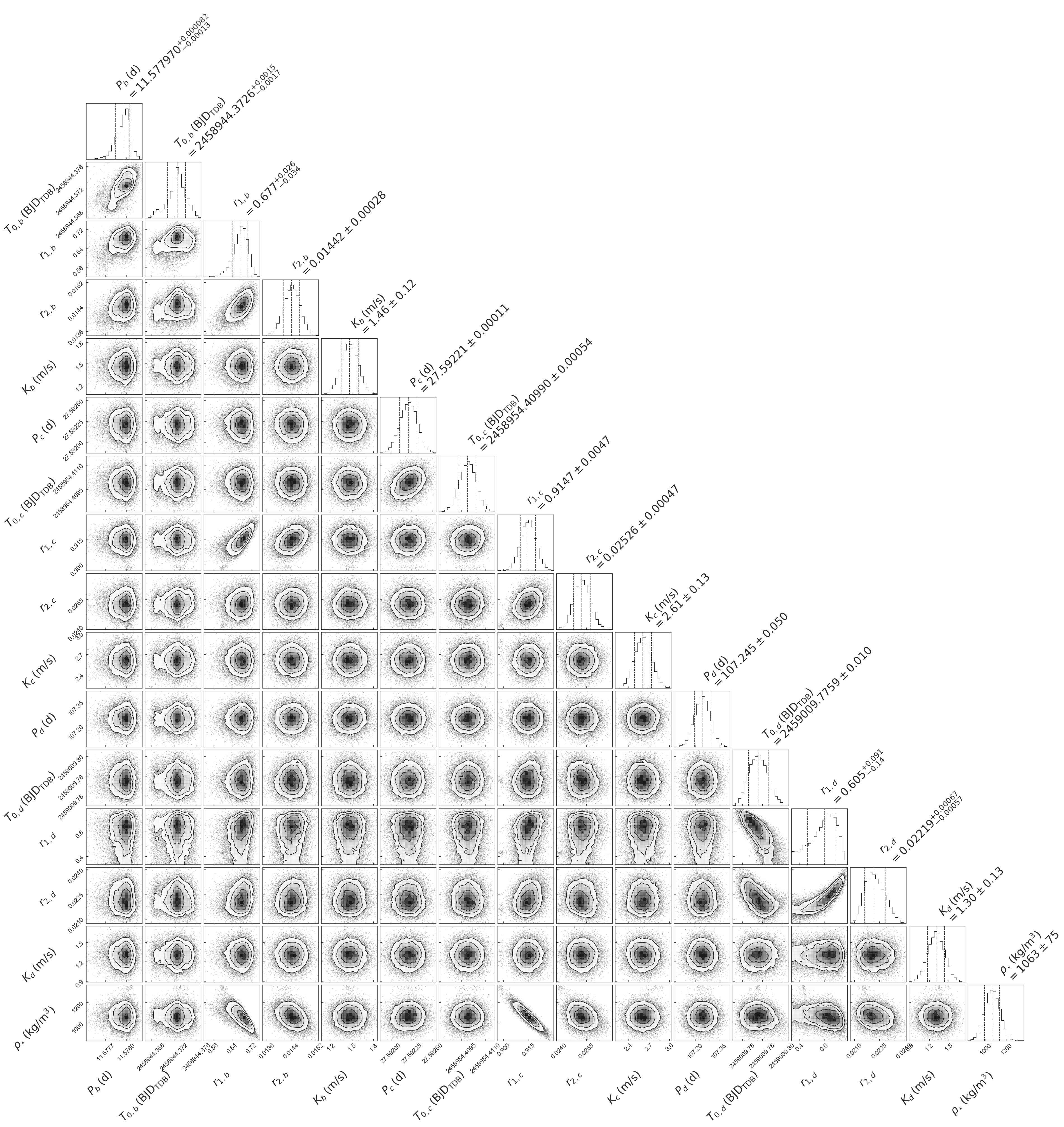}}
\caption{\begin{small}\textbf{Supplementary Figure 7.} Posterior distributions of the main fitted parameters of our global analysis. Shown are the orbital period $P$, the mid-transit time $T_0$, the parameters $r_1$ and $r_2$ (parametrization of the planet-to-star radius ratio $R_p/R_\star$ and the impact parameter $b$), and the RV semi-amplitude $K$ of each planet, as well as the stellar density $\rho_\star$. This plot was made using the \texttt{corner}\cite{corner} python package. \end{small} \label{fig:posterior_all}}
\end{figure*}

\begin{figure*} 
\centerline{\includegraphics[width=0.9\hsize]{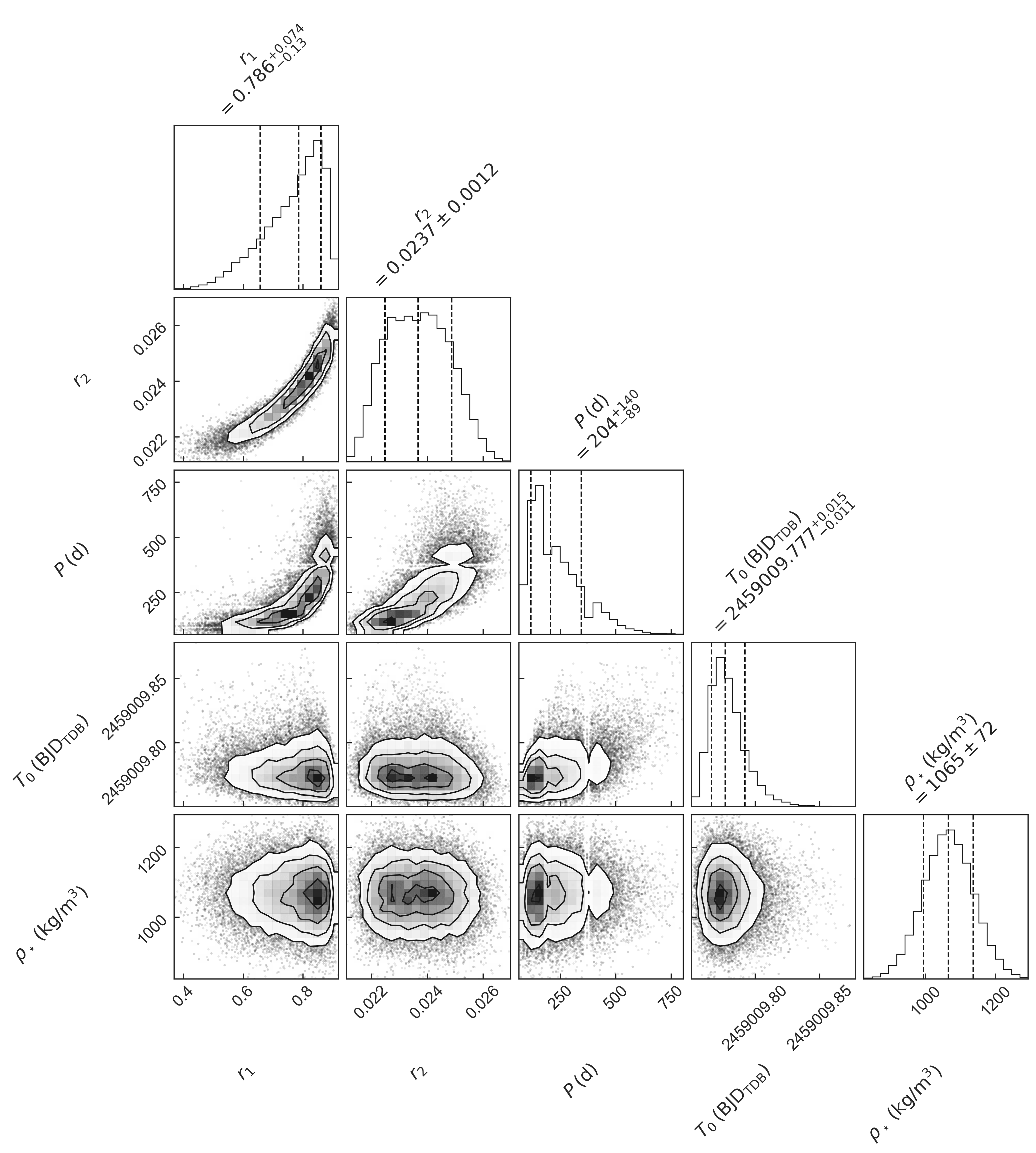}}
\caption{\begin{small}\textbf{Supplementary Figure 8.} Posterior distributions of the transit parameters corresponding to the planet that caused the \textit{CHEOPS} single transit event, as constrained by a combined analysis of all the photometry (\textit{CHEOPS} and \textit{TESS}). The goal here is to assess the constraints brought by the photometry on the orbital period of the transiting object. Shown are the orbital period $P$, the mid-transit time $T_0$, the parameters $r_1$ and $r_2$ (parametrization of the planet-to-star radius ratio $R_p/R_\star$ and the impact parameter $b$), as well as the stellar density $\rho_\star$. This plot was made using the \texttt{corner}\cite{corner} python package. \end{small} \label{fig:corner_3rdplanet}}
\end{figure*}

\begin{figure*}[h!]
\centering
\includegraphics[width=0.8\hsize]{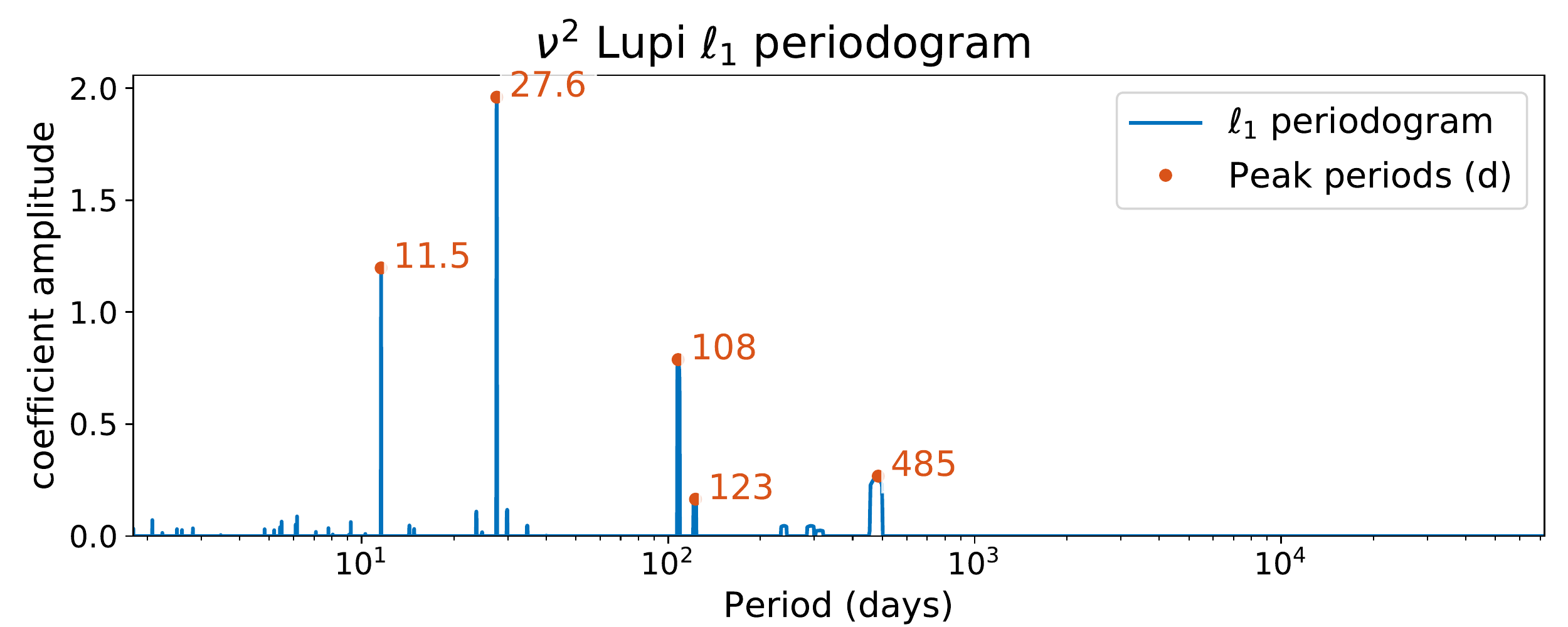}
\hspace{0.5cm}
\includegraphics[width=0.8\hsize]{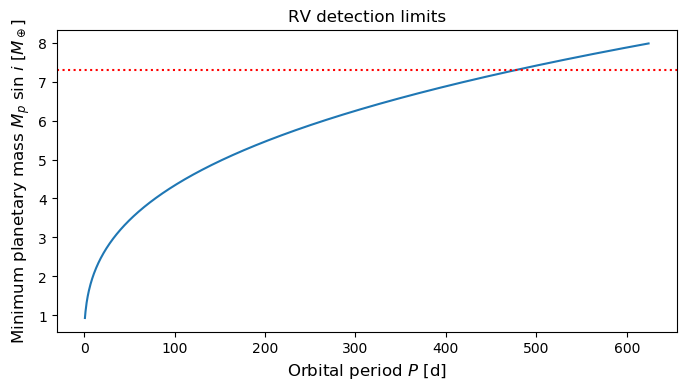}  
\caption{\textbf{Supplementary Figure 9.} \textit{Top:} $\ell_1$-periodogram of the HARPS radial velocities. \textit{Bottom:} Approximate detection limits of the HARPS radial velocities. Any planet located above the blue curve should have been confidently detected. The red dotted line indicates the mass of $\sim$7.3 $M_\oplus$ estimated for the planet that transited during the fifth \textit{CHEOPS} visit, based on its radius (as measured from the transit depth) and the empirical mass-radius relationship of ref. \cite{2017ApJ...834...17C}.}   
\label{fig:RV_4thplanet}
\end{figure*}

\begin{figure*}[h!]
\centerline{\includegraphics[width=14cm,height=8cm]{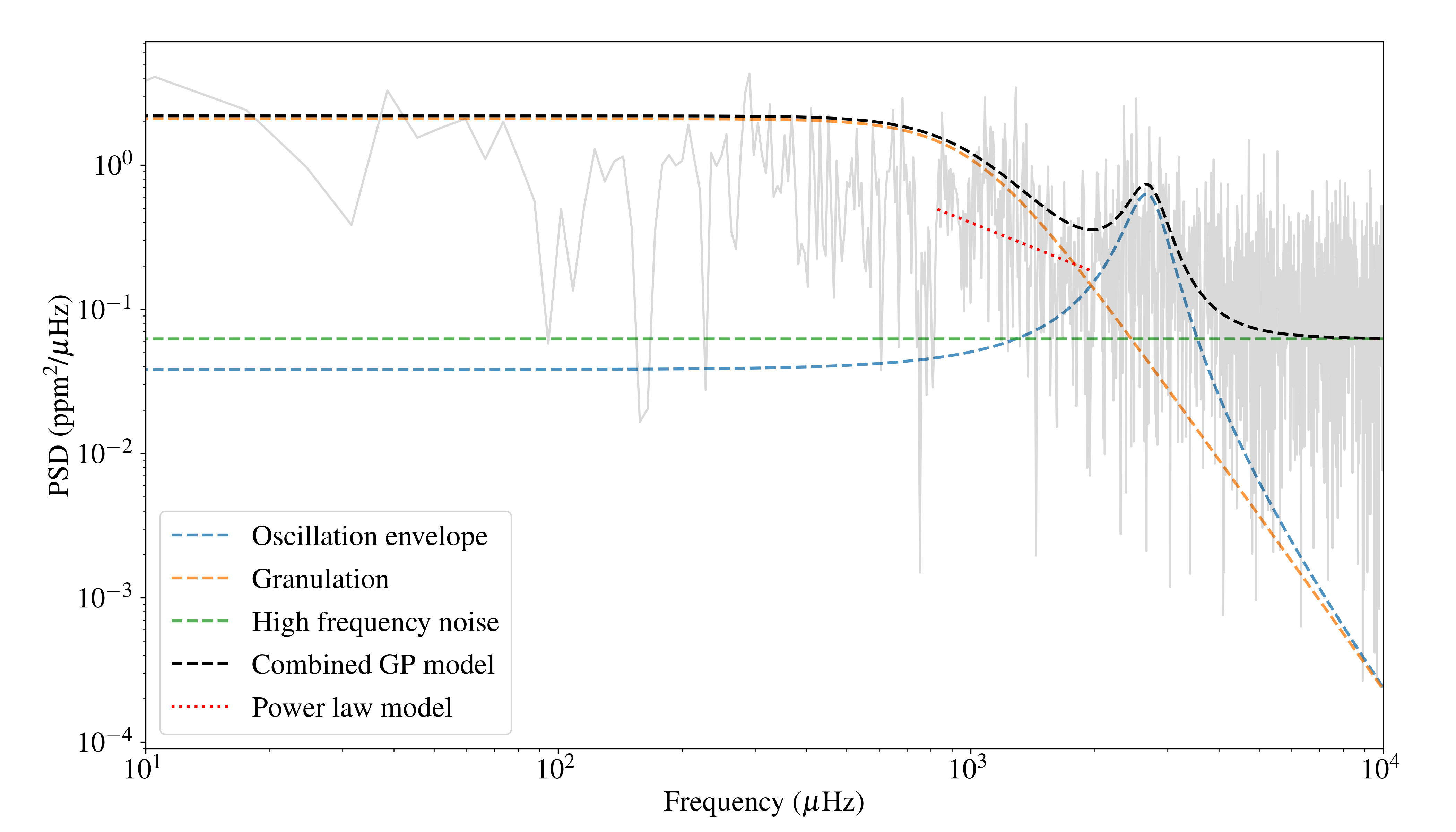}}
\caption{\textbf{Supplementary Figure 10.} Power spectral density (grey) of the photometric residuals obtained after subtracting the best-fit transit and instrumental models from the \textit{CHEOPS} light curves. Dashed lines represent our best-fit models for the different components of the flicker noise, using the GP regression framework of ref. \cite{2019MNRAS.489.5764P}. The red dotted line indicates the slope of the granulation noise when described with a power-law model.\label{fig:CHEOPS_PSD_residuals}}
\end{figure*}

\begin{figure*}[h!]
\centerline{\includegraphics[width=14cm,height=11cm]{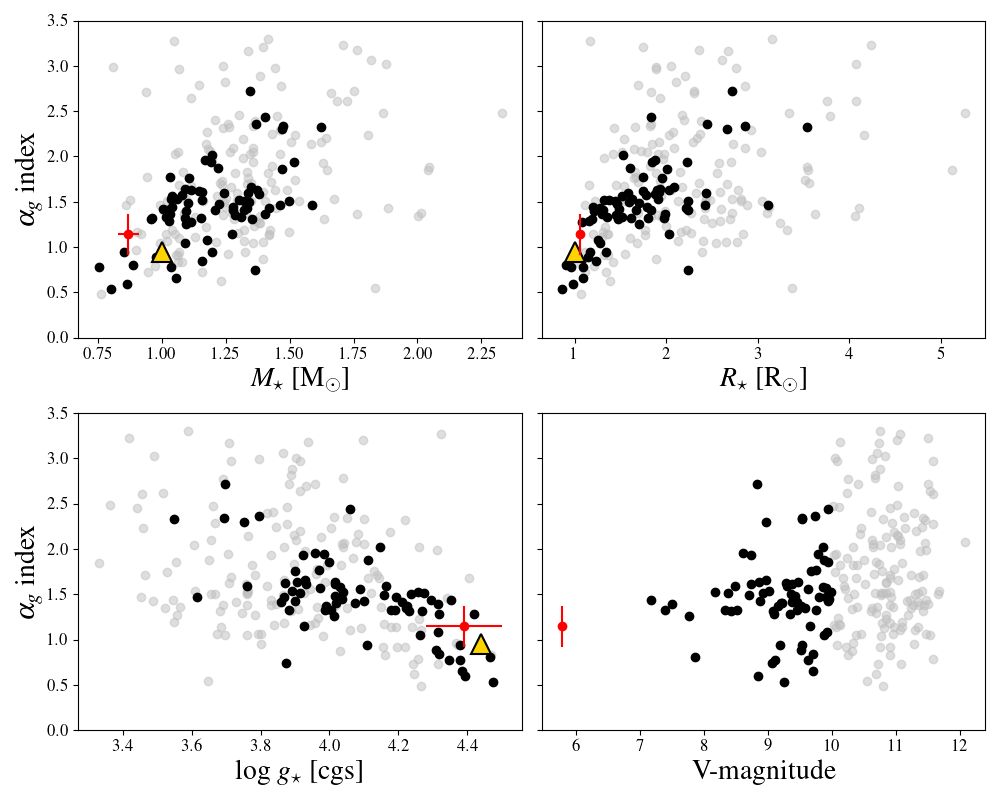}}
\caption{\textbf{Supplementary Figure 11.} \textit{From left to right and top to bottom:} Estimated flicker index associated to granulation as a function of the stellar mass, radius, surface gravity, and apparent $V$-magnitude. Black (resp. grey) dots represent the indices inferred from \textit{Kepler} data (see ref. \cite{2020A&A...636A..70S}) for stars with magnitudes below (resp. above) 10. The yellow triangle represents the flicker index inferred for the Sun from \textit{SOHO/VIRGO} solar data. The red dot with error bars indicates the value obtained for $\nu^2$ Lupi with \textit{CHEOPS}. \label{fig:CHEOPS_flicker_index}}
\end{figure*}

\begin{figure*}[h!]
\centerline{\includegraphics[width=17cm,height=17cm]{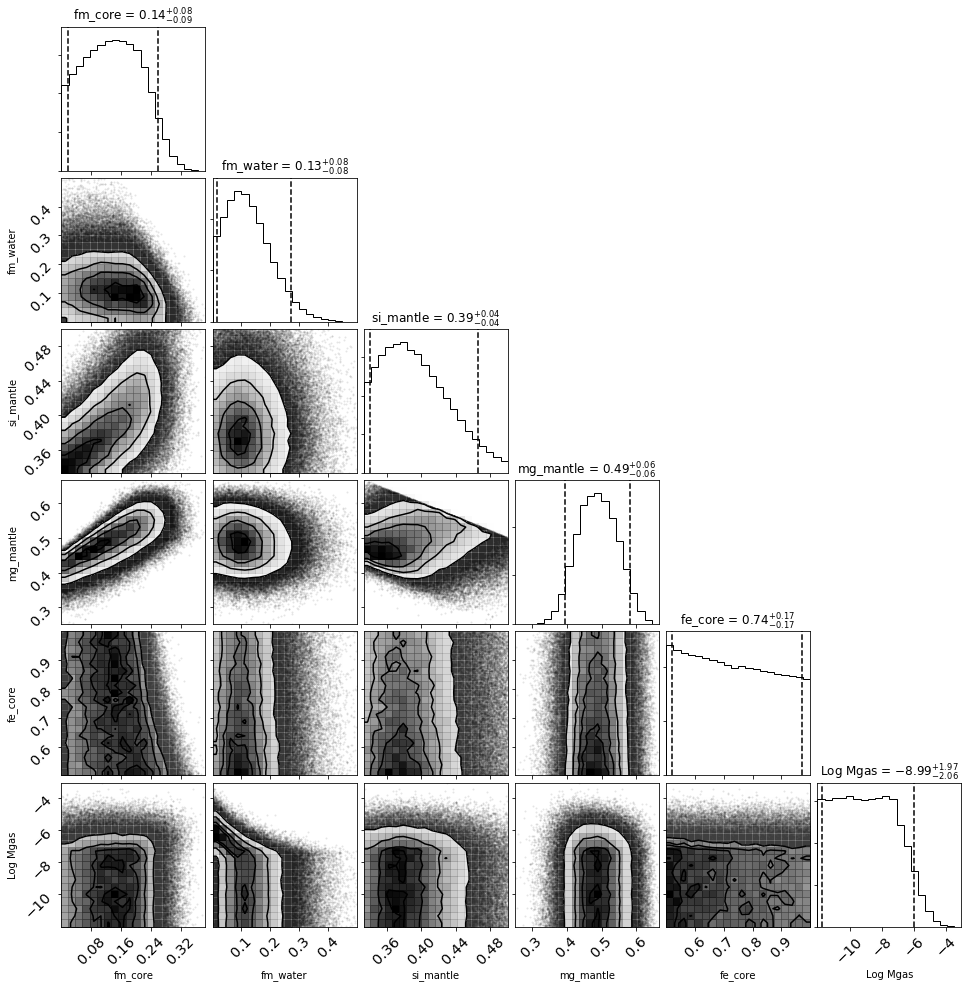}}
\caption{\textbf{Supplementary Figure 12.} Posterior distributions of the main internal structure parameters of planet b. Shown are the mass fractions of the core and water layer (relative to the total amount of heavy elements), the molar fractions of Si and Mg in the mantle, the molar fraction of iron in the core, and the mass of the gas layer. The numbers at the top of each column refer to the mean values and the 5\% and 95\% percentiles.}
\label{fig:internal_structure_b}
\end{figure*}

\begin{figure*}[h!]
\centerline{\includegraphics[width=17cm,height=17cm]{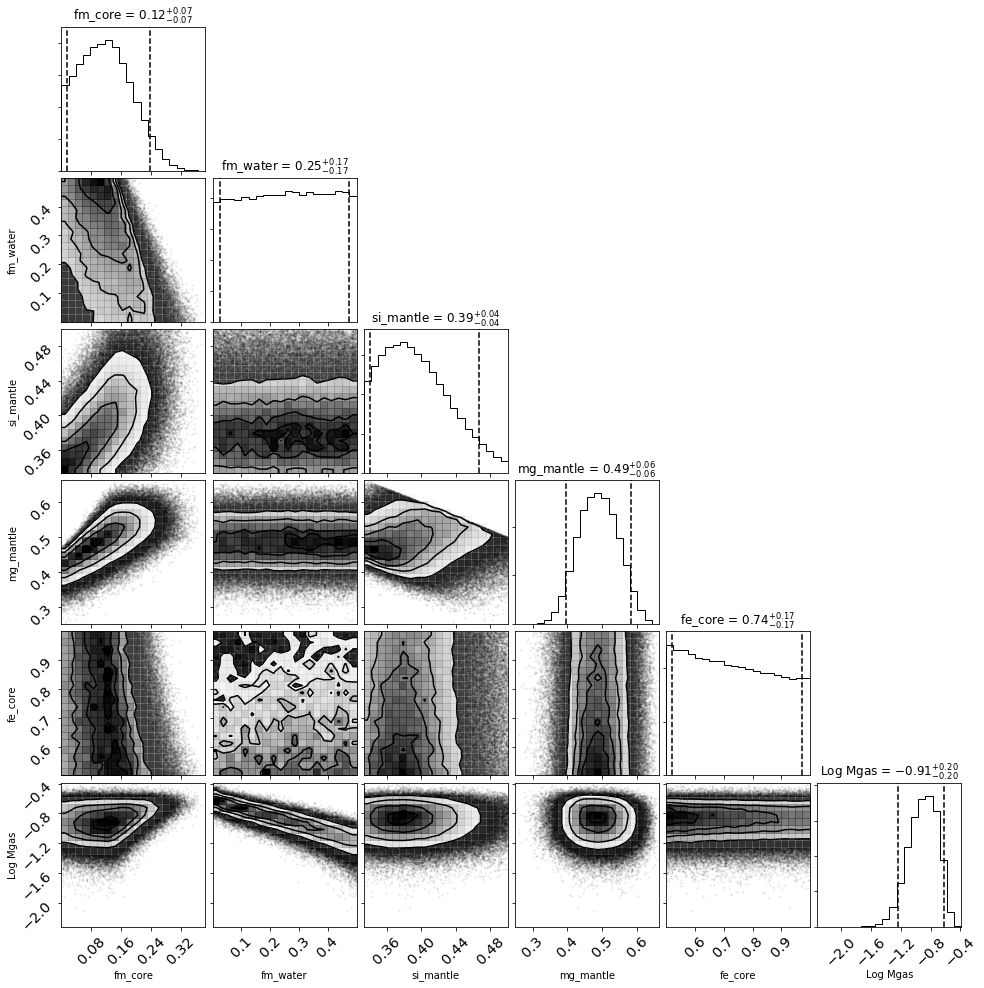}}
\caption{\textbf{Supplementary Figure 13.} Posterior distributions of the main internal structure parameters of planet c. The parameters are the same as in Supplementary Figure \ref{fig:internal_structure_b}.}
\label{fig:internal_structure_c}
\end{figure*}

\begin{figure*}[h!]
\centerline{\includegraphics[width=17cm,height=17cm]{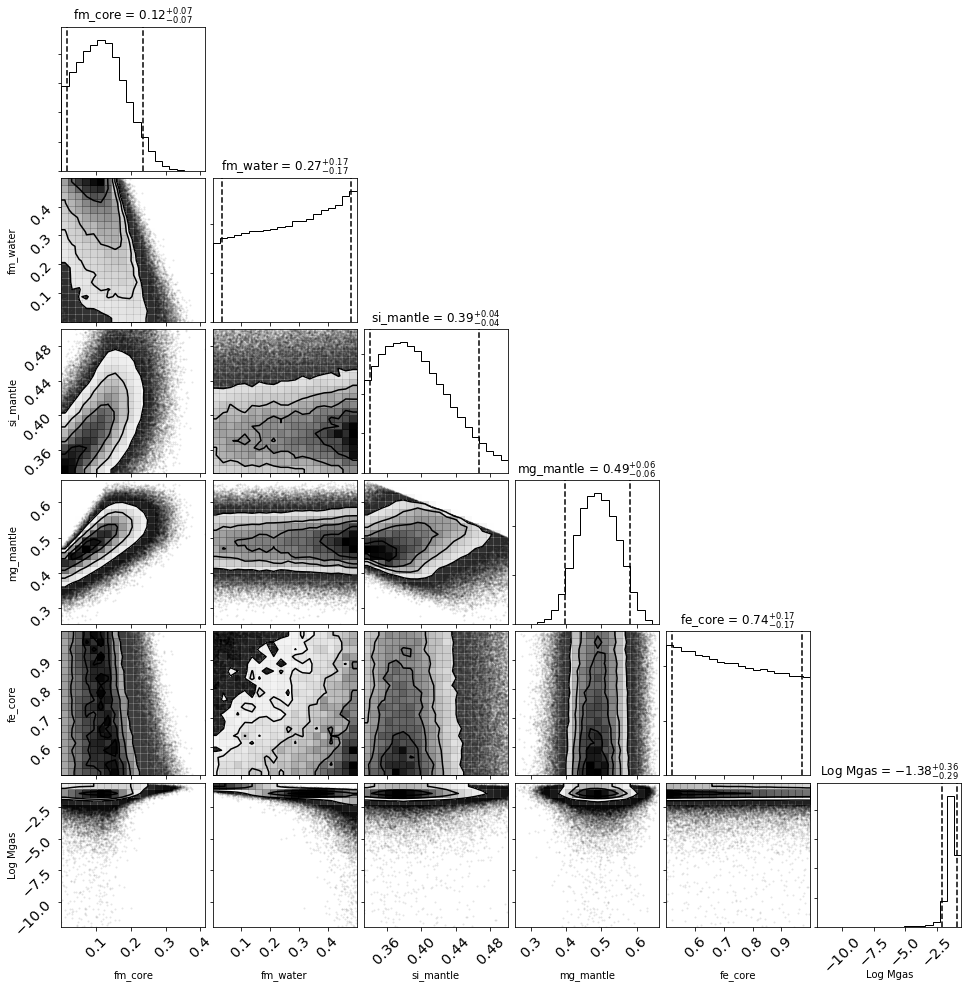}}
\caption{\textbf{Supplementary Figure 14.} Posterior distributions of the main internal structure parameters of planet d. The parameters are the same as in Supplementary Figure \ref{fig:internal_structure_b}.}
\label{fig:internal_structure_d}
\end{figure*}

\begin{figure*}[h!]
\centerline{\includegraphics[width=17cm,height=4cm]{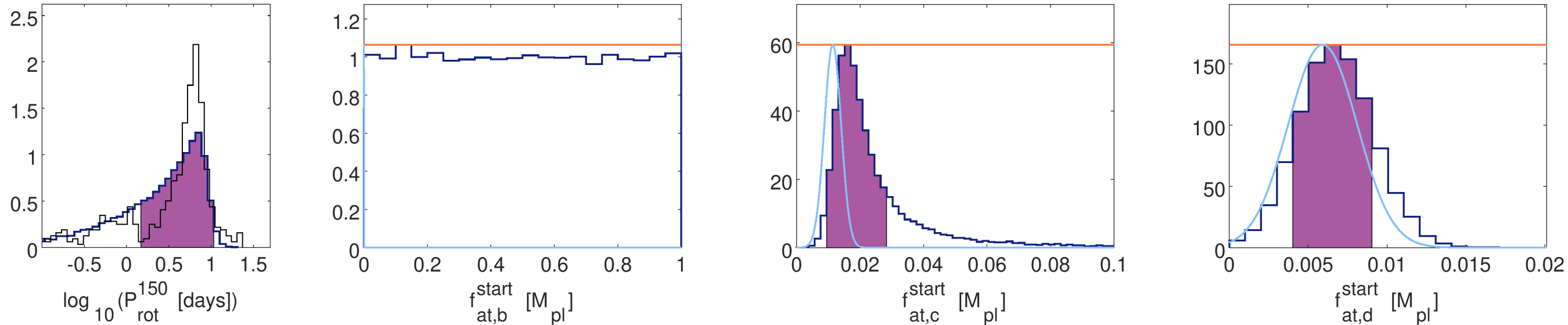}}
\caption{\textbf{Supplementary Figure 15.} \textit{From left to right:} Posterior probability distribution functions (thick dark blue lines) for the stellar rotation period at an age of 150 Myr and for the initial atmospheric mass fractions of \hbox{$\nu^2$ Lupi\,b,} c, and d. The purple areas indicate the 68\% highest posterior density credible interval for each distribution. In the left panel, the distribution marked by the thin black line shows the stellar rotation period distribution\cite{johnstone2015} obtained from measurements of open cluster stars with an age of about 150 Myr and a mass within 0.1 $M_{\odot}$ of \hbox{$\nu^2$ Lupi}'s mass. In the right panels, the horizontal line represents the assumed flat prior arbitrarily shifted vertically to the highest point of each posterior for visualisation purposes. The bright blue line indicates the distribution of the current atmospheric mass fraction returned by our internal structure modelling. Based on this modelling, planet b has basically no gas. Therefore, the bright blue line is not visible in the plot for this planet.}
\label{fig:evolution}
\end{figure*}

\thispagestyle{empty}
\end{document}